\documentclass[prappl,twocolumn,superscriptaddress,showpacs,floatfix]{revtex4-1}
\usepackage[utf8]{inputenc}
\usepackage{graphicx,bm,amssymb,amsmath}
\usepackage{ulem} 
\newcommand{\cN}{{\cal N}}

\begin{document}

\title{Cold-electron bolometer, as a 1~cm wavelength photon counter}
\author{D. V. Anghel}
\affiliation{Horia Hulubei National Institute for R\&D in Physics and Nuclear Engineering, M\u agurele, Romania}
\author{L. S. Kuzmin}
\affiliation{Chalmers University of Technology, Gothenburg 41296, Sweden}
\affiliation{Nizhny Novgorod State Technical University, Nizhny Novgorod 603951, Russia}

\begin{abstract}
We investigate theoretically the possibility of using the cold-electron bolometer (CEB) as a counter for 1~cm wavelength (30~GHz) photons.
To reduce the flux of photons from the environment, which interact with the detector, the bath temperature is assumed to be below 50~mK.
At such temperatures, the time interval between two subsequent photons of 30~GHz  that hit the detector is more than 100 hours, on average, for a frequency window of 1~MHz.
Such temperatures allow the observation of the physically significant photons produced in rare events, like the axions conversion (or Primakoff conversion) in magnetic field.
We present the general formalism for the detector's response and noise, together with numerical calculations for proper experimental setups.
We observe that the current-biased regime is favorable, due to lower noise, and allows for the photons counting at least below 50~mK.
For the experimental setups investigated here, the voltage-biased CEBs may also work as photons counters, but with less accuracy
and, eventually, may require smaller volumes of the normal metal island.
\end{abstract}
\date{\today}

\maketitle

\section{Introduction} \label{sec_intro}

\begin{figure}[t]
  \centering
  \includegraphics[width=9cm,keepaspectratio=true]{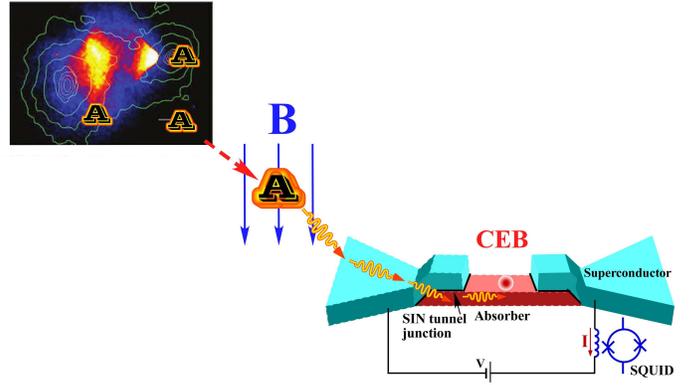}
  \caption{The detection scheme. The photon, produced by the axion in strong magnetic field, is absorbed by the antenna and its energy is dissipated into the normal metal island of the CEB.
  The antenna and the CEB are deposited on an insulating support, not shown here.}
  \label{SINIS_scheme}
\end{figure}

The search for axions~\cite{PhysRevLett.38.1440.1977.Peccei, PhysRevD.16.1791.1977.Peccei, PhysRevLett.40.223.1978.Weinberg, PhysRevLett.40.279.1978.Wilczek} has intensified lately, since they became good candidates for the dark matter in the Universe (see~\cite{PhysDarkUniv.15.2212.2017.Barbieri} and citations therein).
Due to their extremely weak coupling to other massive particles, they are difficult to detect, but they may be converted into photons in intense magnetic fields~\cite{PhysRevLett.51.1415.1983.Sikivie}.
The photons may have low energies (eventually, of the order of 100~$\mu$eV, which correspond to wavelengths of the order of 1~cm) and low flux (one photon in a few hours)~\cite{PhysDarkUniv.15.2212.2017.Barbieri}, so their detection should be attempted with extreme care.
The experimental developments of single photon counters (SPC) that took place over the last decade (see, for example, \cite{ApplPhysLett.96.083505.2010.Santavicca, ApplPhysLett.101.052601.2012.Karasik, PhysRevApplied.3.014007.2015.Gasparinetti}) eventually reached wavelengths of the order of hundreds of microns~\cite{PhysRevLett.117.030802.2016.Govenius}, whereas theoretical estimates show that detection of single-photons of 1~cm wavelength may be in reach~\cite{PhysRevApplied.3.014007.2015.Gasparinetti, PhysRevB.98.205414.2018.Brange}.
For the extreme requirements of axion detection, good candidates for photon counters are devices based on Josephson junctions~\cite{Phys.Lett.1.251.1962.Josephson, Likharev:book}.
Another option is the capacitively coupled cold-electron bolometer (CEB)~\cite{IntWorkshopSupercondNanoElDev.145.2001.Kuzmin, JPhysConfSer.97.012310.2008.Kuzmin, IEEETransApplSupercond.21.3635.2011.Tarasov, ApplPhysLett.86.053505.2005.Schmidt},
which is a symmetric SINIS (superconductor-insulator-normal metal-insulator-superconductor) structure~\cite{ApplPhysLett.68.1996.1996.Leivo}, capacitively coupled to an antenna~\cite{PhysicaB.284-288.2129.2000.Kuzmin, InTech.2012.Kuzmin}, as shown schematically in Fig.~\ref{SINIS_scheme}.
In order to avoid the influence of the strong magnetic field, the detector system is moved away from the sample and the connection between the two parts may be realized by a coaxial cable.
In such a setup, the photon created by the axion decay is captured by the antenna and its energy is dissipated into the central normal metal island, increasing the temperature of the electron gas~\cite{ApplPhysLett.82.293.2003.Anghel}.
This increase in temperature is measured by the SINIS structure, thus the photon is detected.
The two NIS (normal metal-insulator-superconductor) tunnel junctions that form the SINIS structure are used as both, thermometer -- due to the sensitivity of their current-voltage characteristic on the temperature -- and refrigerators~\cite{ApplPhysLett.65.3123.1994.Nahum, ApplPhysLett.68.1996.1996.Leivo} -- to lower the temperature of the electrons in the normal metal island, in order to improve the sensitivity of the detector and decrease its re-equilibration time.
Recently, an important development of the NIS thermometer have been proposed~\cite{PhysRevApplied.10.054048.2018.Karimi}, in which a small gap is induced in the normal metal island by the proximity to a superconductor.
Then, a zero-bias anomaly~\cite{PhysRevLett.17.15.1966.Rowell} appears in the $\cN$IS junction (where $\cN$ stands for the normal metal in which the small gap is induced), which may be used to determine the temperature of the $\cN$ island in a temperature range which may be lower than that accessible to the simple NIS junction.
%
Besides high sensitivity and wide dynamic range, CEBs demonstrate immunity to cosmic rays, due to the tiny volume of the absorber and the decoupling of the phonon and electron subsystems~\cite{JLowTempPhys.176.323.2014.Salatino}.

Counters based on Josephson junctions are under intense investigations (see, for example,~\cite{IEEE_TAS.28.2400505.2018.Kuzmin, PhysRevApplied.8.024022.2017.Walsh, ApplPhysLett.103.142605.2013.Oelsner, PhysRevApplied.7.014012.2017.Oelsner} and citations therein), so in this paper we shall investigate the possibility of using the CEB as a low energy photon counter.

\begin{figure}[t]
  \centering
  \includegraphics[width=9 cm,keepaspectratio=true]{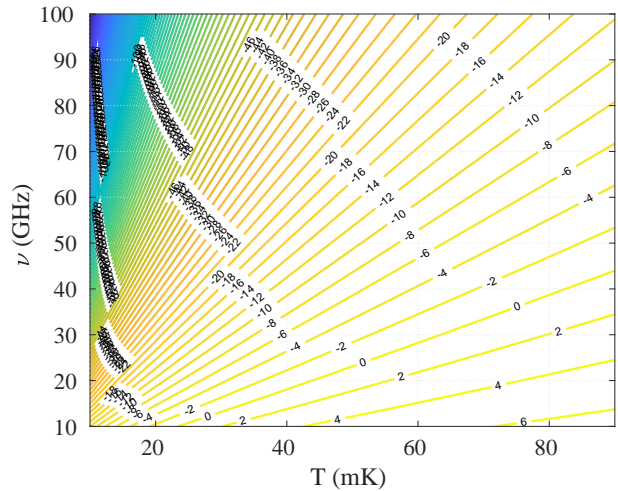}
  \caption{The estimation of the average number of hits on the detector in one hour, by photons from the environment. The area of the antenna is considered to be 1~cm$^2$ and the bandwidth is $\delta\nu=$1~MHz.
  The figure shows the contour plot of $\log_{10} [N_0(\nu) \delta\nu]$.}
  \label{photons_flux}
\end{figure}

In order to be able to identify photons generated by axions decay, the temperature of the environment should be low enough, so that the rate of ``fake events'' (due to photons from the environment hitting the detector) is much smaller than the rate of the ``real events'' (due to photons produced by axions).
If we denote by $N_{ph}(\delta \omega)$ the volume density of the photons from the environment in the narrow frequency window $\delta \omega$, which includes the frequency $\omega$, then
\begin{equation}
  \frac{N_{ph}(\delta\omega)}{\delta\omega} \equiv n_{ph}(\omega) = \frac{1}{\pi^2 c^3} \frac{\omega^2}{e^{\beta\hbar\omega} - 1} , \label{dens_photons}
\end{equation}
where $\beta \equiv 1/(k_BT)$, $k_B$ is the Boltzmann's constant, and $T$ is the bath (environment) temperature--in Eq.~(\ref{dens_photons}) we took into account the two photon polarizations.
From Eq.~(\ref{dens_photons}) we obtain the flux of photons on the unit area of the detector's surface,
\begin{equation}
  \phi (\omega) \equiv \int_0^{2\pi} d\phi \int_0^{\pi/2} d\theta\, \sin\theta \frac{c n_{ph}(\omega)}{4\pi}
  = \frac{1}{2 \pi^2 c^2} \frac{\omega^2}{e^{\beta\hbar\omega} - 1} . \label{flux}
\end{equation}
For photons of wavelength of the order of 1~cm (30~GHz frequency), the area of the detector plus antenna is of the order of $A_0 = 1~{\rm cm}^2$.
The estimation of the average number of photon hits on the detector in the time interval $t_0 = 1$~hour and in a unit frequency window of 1~Hz is $N_0(\nu) = A_0 t_0 \phi(2\pi\nu)$, where $\nu \equiv \omega/(2\pi)$.
In Fig.~\ref{photons_flux} we show the contour plot of $\log_{10}[N_0(\nu) \delta\nu]$, for a typical bandwidth of $\delta \nu = 1$~MHz.
We observe that for photons of 30~GHz we have, on average, one photon hit in more than 100 hours, if the temperature of the environment is 50~mK.
This allows enough room for an accurate detection of photons generated by axions decay.

The paper is organized as follows.
In the next section we analyze the response of the CEB, namely the temperature increase due to the photon absorption, the signal produced in the measured quantity (current or voltage), and the re-equilibration time (the time scale in which the device cools down, after the photon absorption).
In Section~\ref{sec_noise} we calculate the noise in the system, to see if the signal produced by the photon can be observed.
In Section~\ref{sec_concl} we draw the conclusions.

While in the main body of the paper, the calculations are done for a volume of the normal metal $\Omega = 0.01~{\rm \mu m}^3$, in the Appendix we present the main results for $\Omega = 0.1~{\rm \mu m}^3$, to emphasize the flexibility that exists in the construction of the device and its limitations.

\section{Response of the cold-electron bolometer} \label{sec_response}

The principle of detection is presented in Fig.~\ref{SINIS_scheme} (see, for  example, Refs.~\cite{ApplPhysLett.82.293.2003.Anghel, ApplPhysLett.82.3970.2003.Luukanen, ApplPhysLett.68.1996.1996.Leivo, ApplSupercond.5.227.1998.Leivo, RevModPhys.78.217.2006.Giazotto}).
The superconducting antenna is coupled to the normal metal island by two NIS tunnel junctions, forming the symmetric SINIS structure.
The whole detector is deposited on an insulating support (not shown in the figure).
The thicknesses of the metallic layers (normal metal and superconductor) are 10 to 20~nm.
When a photon, absorbed in the antenna, dissipates its energy into the normal metal island, the heat diffuses in the normal metal in the characteristic time $\tau_d \approx L^2 / (\pi^2 D)$, where $L$ is the linear dimension of the normal metal and $D$ is the electrons' diffusion constant.
For typical values, $L \sim 1~{\rm \mu m}$ and $D = 10^{-4}~{\rm m^2 s^{-1}}$, we obtain $\tau_d \sim 1$~ns~\cite{ApplPhysLett.82.293.2003.Anghel}, which sets the lower limit for the detection time.

At working temperatures, which are below 100~mK, the electron system in the normal metal is very weakly coupled to the phonons system~\cite{PhysRevB.49.5942.1994.Wellstood, SolidStateCommun.227.56.2016.Anghel, PhysRevB.93.115405.2016.Cojocaru, EurPhysJB.90.260.2017.Anghel, PhysScr.94.105704.2019.Anghel}.
This allows for the independent thermalization of the electrons system, at temperature $T_e$, and of the phonons system, at temperature $T_{ph}$.
Then, the heat power between the electrons system and phonons system may be written in general as
\begin{equation}
  \dot Q_{ep} = \Sigma_{ep} \Omega \left(T_e^x - T_{ph}^x \right) , \label{Qep}
\end{equation}
where $\Sigma_{ep}$ is the coupling constant, $\Omega$ is the volume of the normal metal, and the exponent $x$ depends on the model and the dimensionality of the phonons system (in our case, $x$ may take values between 3.5 and 5)~\cite{PhysRevB.49.5942.1994.Wellstood, PhysRevB.81.245404.2010.Viljas, SolidStateCommun.227.56.2016.Anghel, PhysRevB.93.115405.2016.Cojocaru, EurPhysJB.90.260.2017.Anghel}.
Since the heat power exchanged between electrons and phonons is low, we shall assume in general that $T_{ph}$ is equal to the heat bath temperature $T_b$.

In the absence of photons, the electrons equilibrate at temperature $T_{e1}$, determined by the balance between the heat exchanged with the phonons (Eq.~\ref{Qep}) and the heat extracted from the normal metal into the superconductor, through the two NIS junctions (Eq.~\ref{def_QJ} below)--in addition to these, spurious power injection may be present, as observed in Ref.~\cite{PhysRevApplied.3.014007.2015.Gasparinetti}, but in the absence of a quantitative understanding, we do not take it into account.
When a photon is absorbed, its energy is dissipated into the electrons system of the normal metal, increasing its temperature to $T_{e2}$.
In the low temperature limit, the internal energy of the electron system is, in general, proportional to $T_e^2$~\cite{Stone:book} and to the volume, so we may write
\begin{equation}
  U (T_e) = \Omega C_e T_e^2 . \label{lowT_U2}
\end{equation}
For concrete systems, $C_e$ is either a fitting parameter or is determined by the theoretical model.
For simplicity, we take the value corresponding to an ideal gas, namely $C_e = (2 m_{\rm e}/\hbar^2)^{3/2} \epsilon_F^{1/2} k_B^2/12$, where $\epsilon_F$ is the Fermi energy of the electrons, $m_e$ is the electron's mass, $k_B$ is the Boltzmann's constant, and $\hbar$ is the reduced Plank's constant.
Nevertheless, significant differences may appear between the calculated values of $C_e$ and the measured ones~\cite{PhysRevApplied.3.014007.2015.Gasparinetti}.
If we denote by $\omega_{ph}$ the angular frequency of the photon, using Eq.~(\ref{lowT_U2}) we can write the equation
\begin{equation}
  \hbar \omega_{ph} \equiv \epsilon_{ph} = U (T_{e2}) - U (T_{e1}) , \label{temp_var}
\end{equation}
where $\epsilon_{ph}$ is the energy of the photon.

The heat and charge transport through the NIS junctions have been extensively studied in the past (for example, see Refs.~\cite{ApplPhysLett.63.3075.1993.Nahum, ApplPhysLett.65.3123.1994.Nahum, ApplPhysLett.66.3203.1995.Nahum, ApplPhysLett.68.1996.1996.Leivo, ApplSupercond.5.227.1998.Leivo, RevModPhys.78.217.2006.Giazotto,JLowTempPhys.93.733.1993.Nahum, JApplPhys.83.3217.1998.Jochum, JLowTempPhys.123.197.2001.Anghel}).
We shall assume that the junctions are identical, of normal resistance $R_T$, and they are biased with opposite voltages, $V$ and $-V$.
If the tunneling resistance is big enough, Andreev reflection does not occur and particles and energy are transported by quasiparticle tunneling.
We define, like in Refs.~\cite{JApplPhys.83.3217.1998.Jochum, JLowTempPhys.123.197.2001.Anghel}, four tunneling currents,
\begin{subequations} \label{tunneling_js}
\begin{eqnarray}
  j_1(\epsilon) &\equiv& \frac{g(\epsilon)}{e^2 R_T} f (\epsilon - eV, T_e) [1 - f(\epsilon, T_s)] \label{tunneling_j1} \\
  j_2(\epsilon) &\equiv& \frac{g(\epsilon)}{e^2 R_T} f (\epsilon + eV, T_e) [1 - f(\epsilon, T_s)] \label{tunneling_j2} \\
  j_3(\epsilon) &\equiv& \frac{g(\epsilon)}{e^2 R_T} [1 - f (\epsilon - eV, T_e)] f(\epsilon, T_s) \label{tunneling_j3} \\
  j_4(\epsilon) &\equiv& \frac{g(\epsilon)}{e^2 R_T} [1 - f (\epsilon + eV, T_e)] f(\epsilon, T_s) \label{tunneling_j4}
\end{eqnarray}
\end{subequations}
where $\Delta$ and $\epsilon (\ge \Delta)$ are the energy gap and the quasiparticle energy in the superconductor, respectively, whereas $g(\epsilon) \equiv \epsilon / \sqrt{\epsilon^2 - \Delta^2}$ is proportional to the quasiparticle density of states
(for a detailed description of the currents~\ref{tunneling_js}, see~\cite{JApplPhys.83.3217.1998.Jochum}).
By $T_s$ we denoted the quasiparticles' temperature in the superconductor and, because of the low heat power exchanged between the normal metal and the superconductor, we assume that $T_s = T_b$.
The density of states $g(\epsilon)$ may be generalized, to include sub-gap tunneling~\cite{PhysRevLett.105.026803.2010.Pekola, PhysRevB.88.174507.2013.DiMarco}, but since this is determined by the interaction with the environment, we neglect it in this analysis (like, for example, in~\cite{PhysRevApplied.3.014007.2015.Gasparinetti, PhysRevB.98.205414.2018.Brange}).
Using Eqs.~(\ref{tunneling_js}), the electrical current and the heat current through an NIS junction are~\cite{JApplPhys.83.3217.1998.Jochum, JLowTempPhys.123.197.2001.Anghel, JApplPhys.89.6464.2001.Golubev}
\begin{subequations} \label{defs_IJ_QJ}
\begin{eqnarray}
  I_J &=& e \int_\Delta^\infty (j_1 - j_2 - j_3 + j_4) d\epsilon \label{def_IJ} \\
  &=& \frac{1}{e R_T} \int_\Delta^\infty g(\epsilon) [f (\epsilon - eV, T_e) - f (\epsilon + eV, T_e)] d\epsilon \nonumber
\end{eqnarray}
and
\begin{equation}
  \dot Q_J = \int_\Delta^\infty [(\epsilon - eV)(j_1 - j_3) - (\epsilon + eV) (j_4 - j_2)] d\epsilon , \label{def_QJ}
\end{equation}
\end{subequations}
respectively.
$I_J$ and $\dot Q_J$ are positive when the current and heat, respectively, flows from the electrons of the normal metal into the superconductor.
The equilibrium temperature of the electron system is obtain by equating the total power $\dot Q_T(T_e, T_b) \equiv \dot Q_{ep}(T_e, T_b) + 2 \dot Q_J(T_e, T_b)$) to zero (the factor 2 appears because of the two NIS junctions attached to the normal metal), namely
\begin{equation}
  \dot Q_T (T_{e}, T_b) = 0 . \label{power_bal}
\end{equation}
Equation~(\ref{power_bal}) represents the \textit{heat balance equation} for our system.

\begin{figure}[t]
  \centering
  \includegraphics[width=9cm]{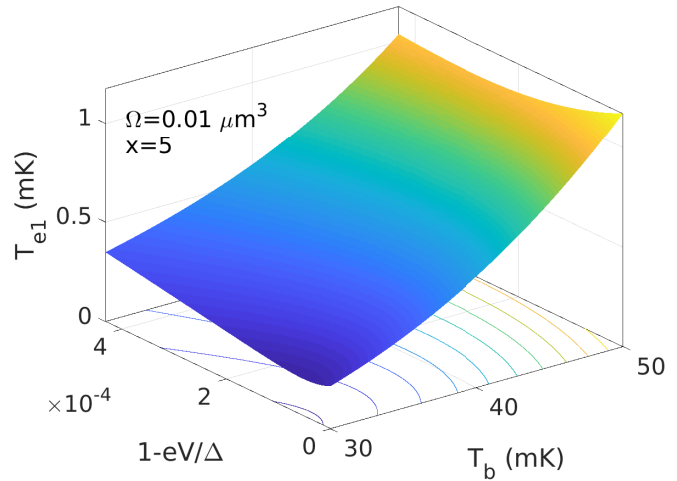}
  \caption{The theoretical estimation of the equilibrium temperature of the (ideal) electron gas in the normal metal island, cooled by the SINIS structure, starting from the bath temperature $T_b$ and for the bias voltage $V$ (for each NIS junction). The tunneling resistance of each of the two junctions is $R_T = 35~{\rm k\Omega}$ and the volume of the normal metal island is $0.01~{\rm \mu m}^3$.
  }
  \label{T_equil_RT35kO_Vol0o01microm3_nu30GHz}
\end{figure}

We consider that the normal metal is Cu and the superconductor is Al.
In Fig.~\ref{T_equil_RT35kO_Vol0o01microm3_nu30GHz} we plot the solutions of Eq.~(\ref{power_bal}), for $\dot Q_{ep}$ given by Eq.~(\ref{Qep}), with $x = 5$ and $\Sigma_{ep} = 4\times10^9~{\rm W m^{-3} K^{-5}}$~\cite{ApplPhysLett.68.1996.1996.Leivo}.
The tunneling resistance is $R_T = 35~{\rm k\Omega}$ for each of the two NIS junctions, the volume $\Omega = 0.01~{\rm \mu m}^3$, and the energy gap in the superconducting Al is $\Delta = 0.2$~meV~\cite{ApplPhysLett.68.1996.1996.Leivo, APL76.2782.2000.Pekola}.
We shall use these numerical values throughout the paper, except for the Appendix.
To emphasize the dependence of these results on the volume $\Omega$, in the Appendix we shall plot the results of the same calculations, but with $\Omega = 0.1~{\rm \mu m}^3$.
We use the experimental values for $x$ and $\Sigma_{ep}$, instead of the theoretical ones, calculated more recently, because the values existing in the literature (both, theoretical and experimental) vary significantly from model to model and from sample to sample (see the diversity of results presented, for example, in Refs.~\cite{PhysRevB.49.5942.1994.Wellstood, ApplPhysLett.68.1996.1996.Leivo, Stroscio_Dutta:book, PRL.99.145503.2007.Karvonen,JPhysConfSer.92.012043.2007.Karvonen, PhysRevLett.68.1156.1992.DiTusa, PhysRevB.72.224301.2005.Qu, PhysRevB.51.9930.1995.Bannov, PhysRevB.77.033401.2008.Hekking, SolidStateCommun.227.56.2016.Anghel, PhysRevB.93.115405.2016.Cojocaru}).


The temperature increase $\Delta T_e \equiv T_{e2} - T_{e1}$ due to the absorption of a 1~cm wavelength photon leads to an increase of the current (at voltage-bias) or a decrease of the voltage (at current-bias).
The photon can be detected if the variation of the measured quantity, $I_J$ or $V$, is bigger than the noise (i.e. the mean square fluctuation of the measured quantity) that we shall calculate in Section~\ref{sec_noise}.

\subsection{Detector re-equilibration}

\begin{figure}[t]
  \centering
  \includegraphics[width=9cm,keepaspectratio=true]{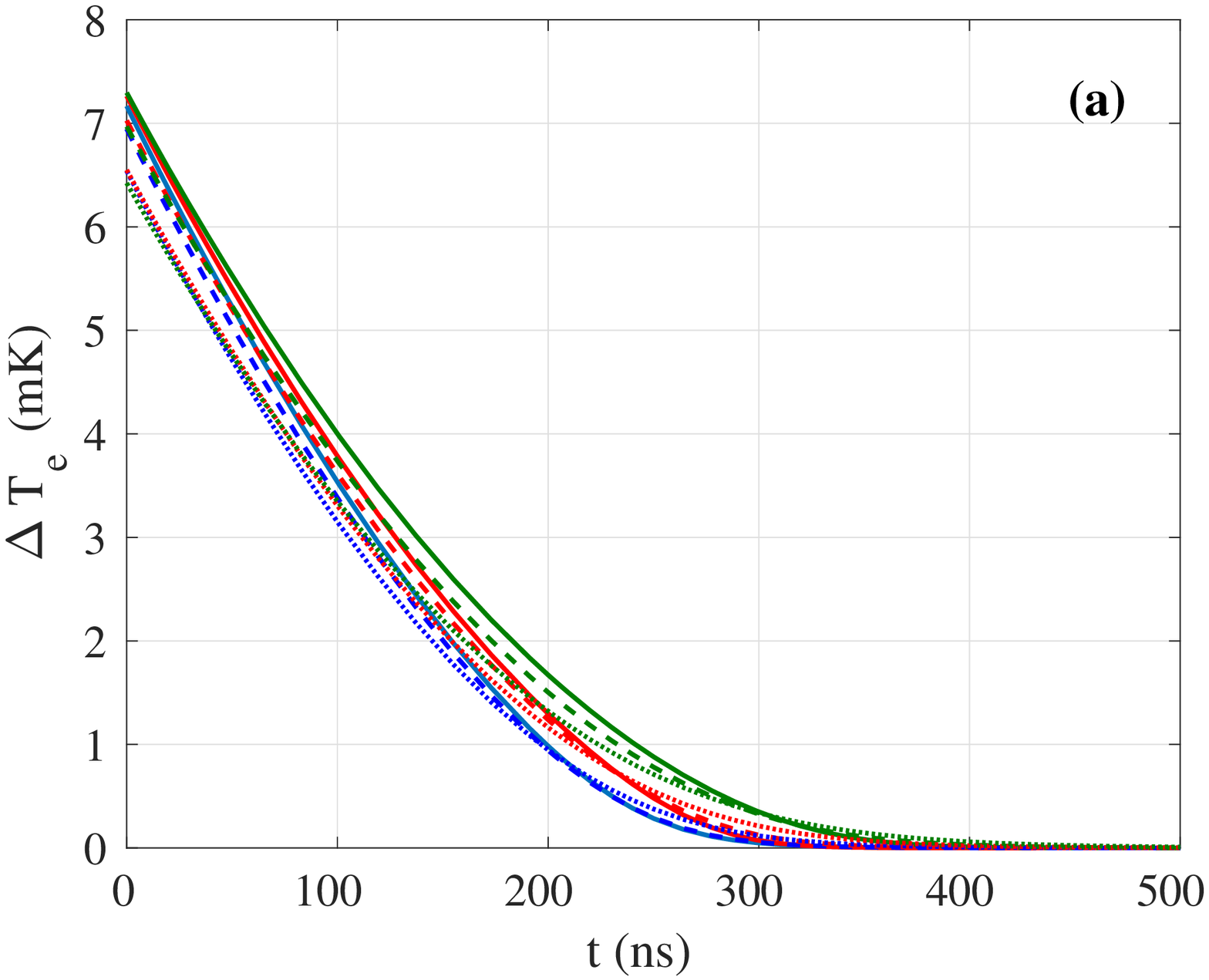}
  \includegraphics[width=9cm,keepaspectratio=true]{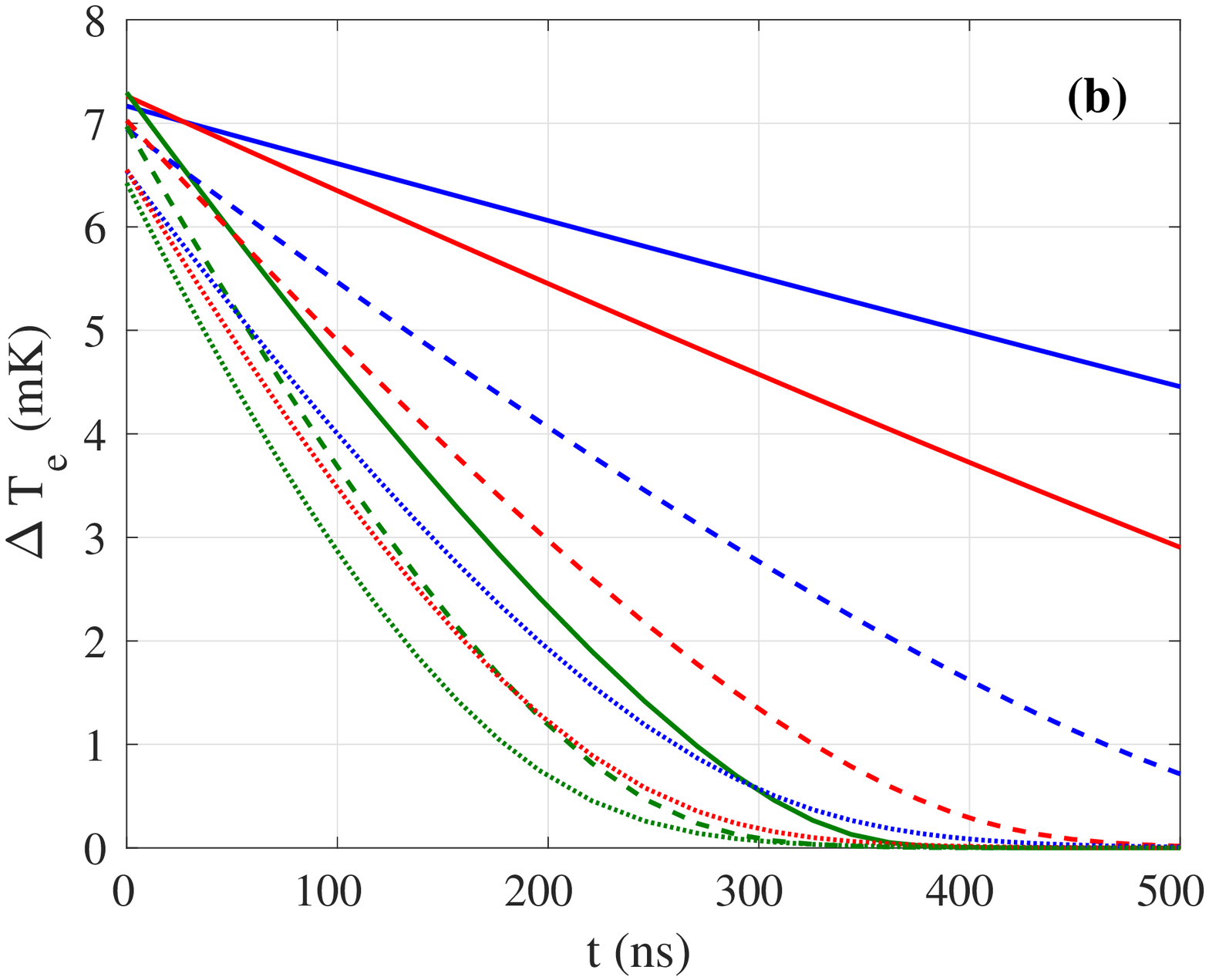}
  \caption{(Color online) The time evolution of the electron temperature in the normal metal at voltage- (a) and current-bias (b).
  The volume is $\Omega = 0.01~{\rm \mu m}^3$.
  The  bath temperature takes three values: 30~mK (solid lines), 40~mK (dashed lines), and 50~mK (dotted lines).
  In (a), the voltages are fixed at $V_0$, with $1-eV_0/\Delta_0 = 0.43\times 10^{-3}$ (blue lines), $0.22\times 10^{-3}$ (red lines), and 0 (green lines).
  In (b) the currents are fixed $I_J(T_{e1}, V_0)$, where $V_0$ correspond to the same values as in (a).
  }
  \label{time_evol_Te}
\end{figure}

We calculate the re-equilibration time of the detector, $\tau$, which sets the time scale in which the temperature returns to the initial value after the photon absorption.
Let's say that at time $t=0$, the temperature of the electrons in the normal metal is varied by $\Delta T_e(0) \equiv T_{e2} - T_{e1}$, after which they cool back to $T_{e1}$.
If $\Delta T_e (0) \ll T_{e1}$, then we can assume an exponential time dependence,
\begin{subequations} \label{defs_taus}
\begin{equation}
  \Delta T_e (t) = \Delta T_e(0) e^{-t/\tau}, \label{Te_f_t}
\end{equation}
where, from the expression of $\dot Q_T$, we obtain
\begin{equation}
  \tau^{-1} = \frac{1}{C_V} \left( \frac{\partial \dot Q_T}{\partial T_e} \right) \equiv \tau_J^{-1} + \tau_{ep}^{-1} \ll \tau_d^{-1} ,
\end{equation}
in obvious notations.
From Eqs.~(\ref{def_QJ}) and (\ref{Qep}) we obtain
\begin{equation}
  \frac{1}{\tau_J} = \frac{2}{C_V} \left( \frac{\partial \dot Q_J}{\partial T_e} \right) \quad {\rm and} \quad
  %
  \frac{1}{\tau_{ep}} = \frac{5 \Sigma_{ep} \Omega T_e^4}{C_V} , \label{tau_ep}
\end{equation}
\end{subequations}
where $C_V \equiv C_V (T_e)$ is the heat capacity of the electron system in the normal metal and $\partial \dot Q_J/\partial (k_BT_e)$ is calculated in~(\ref{dQJ_deV}).
\begin{figure}[t]
  \centering
  \includegraphics[width=9cm,keepaspectratio=true]{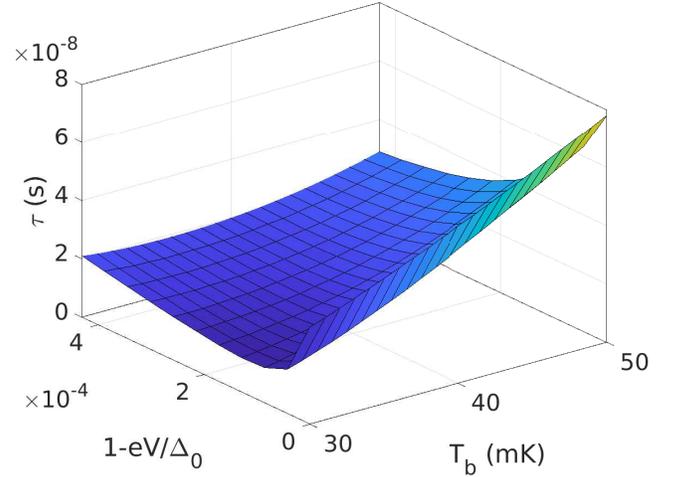}
  \caption{The relaxation time $\tau$.}
  \label{relaxation_time}
\end{figure}
The dependence of $\tau$ on the bath temperature and bias voltage is plotted in Fig.~\ref{relaxation_time}.
Nevertheless, if $\Delta T_e$ is comparable or bigger than $T_{e1}$ -- as we shall see it happens in our case -- Eqs.~(\ref{defs_taus}) give only the order of magnitude of the time required for the detector to re-equilibrate.
The time variation of the temperature in the general case is given by the formula
\begin{eqnarray}
  C_V(T_e) \frac{d T_e}{dt} = \dot Q_T (T_e, T_b) . \label{dTe_dt}
\end{eqnarray}
In Fig.~\ref{time_evol_Te} we show the time evolution of the temperature difference $\Delta T_e (t)$ for a few representative values of $T_b$ and bias voltages or currents.
We observe that, in accordance to the results plotted in Fig.~\ref{relaxation_time}, the relaxation time is of the order of a few tens of nanoseconds.
We also observe that the relaxation time may depend strongly on the type of bias used (current- or voltage-bias), especially at low temperatures.

\section{Noise} \label{sec_noise}

Since we analyze only the CEB, in the noise calculations we neglect the fluctuations produced in the external circuit.
These depend on the experimental setup and should be added to the the fluctuations calculated here.

For integrating detectors (that is, detectors that measure the total incoming radiation flux) the figure of merit is the \textit{noise equivalent power} (NEP).
If $M$ is the measured quantity for such a detector (e.g. the current, in a voltage biased CEB), the NEP represents the input radiation power on the unit bandwidth $\dot Q_{oe}(\omega)$ that produces a signal equal in amplitude to the square root of the spectral density of noise, $(\langle |\delta M(\omega)|^2 \rangle)^{1/2}$. Concretely,
\begin{equation}
  {\rm NEP} \equiv \frac{(\langle |\delta M(\omega)|^2 \rangle)^{1/2}}{|\partial M / \partial \dot Q_{oe} |} . \label{def_NEP0}
\end{equation}
In the voltage-biased CEB, the measured quantity is the current and the spectral density of its fluctuation is~\cite{ApplPhysLett.82.293.2003.Anghel, JLowTempPhys.123.197.2001.Anghel}
\begin{subequations} \label{tot_fluct_IJ_VJ_omega}
\begin{eqnarray}
  \langle |\delta I_{J} (\omega)|^2 \rangle &=& \langle |\delta I_{J {\rm shot}} (\omega)|^2 \rangle + \left( \frac{\partial I_J}{\partial T_e} \right)^2 \langle |\delta T_e(\omega)|^2 \rangle \nonumber \\
  && + \frac{1}{e^2 \omega^2} \left( \frac{\partial \epsilon_F}{\partial N} \frac{\partial I_J}{\partial (eV)} \right)^2 \langle |\delta I_J (\omega)|^2 \rangle \nonumber \\
  && + 2 \frac{\partial I_J}{\partial T_e} \Re \left( \langle \delta I_{J}(\omega) \delta T^*_e(\omega) \rangle \right) \label{tot_fluct_IJ_omega} \\
  && + \frac{2}{e} \frac{\partial \epsilon_F}{\partial N} \frac{\partial I_J}{\partial (eV)} \frac{\partial I_J}{\partial T_e} \Re \left( \left\langle \delta T^*_e (\omega) \frac{\delta I_J (\omega)}{i \omega} \right\rangle \right) , \nonumber
\end{eqnarray}
where the angular brackets $\langle \cdot \rangle$ denote averages, $\Re (\cdot)$ represents the real part of a complex number, $\delta I_J (\omega)$ and $\delta T_e(\omega)$ are the Fourier transforms of the noise in current and temperature, respectively, $\delta T^*_e$ is the complex conjugate of $\delta T_e$, whereas $\delta I_{J,{\rm shot}}(\omega)$ is the Fourier transform of the current shot noise (for details, see Appendix~\ref{sec_current_fluct}).
The first term on the right hand side of Eq.~(\ref{tot_fluct_IJ_omega}), namely $\delta I_{J,{\rm shot}}(\omega)$, represents the lowest order contribution to the noise.

In  the current biased setup, the voltage fluctuation is ultimately determined by the fluctuation of the number of electrons in the normal metal island $N$ (since the positive charge does not change).
The charging on the normal metal island leads to a change of its potential energy, which, in turn, influences the charges on the junctions capacitances--therefore, the  voltage on the CEB.
These processes may be studied in more detail, but
for a CEB the capacitance of the normal metal island is big enough, so we use the linear approximation assuming that the voltage fluctuations are proportional to the fluctuations of the number of electrons in the normal metal island.
Concretely, we use $\delta (eV) = (\partial \epsilon_F/\partial N) \delta N$ and write the spectral density of the voltage fluctuation as~\cite{ApplPhysLett.82.293.2003.Anghel}
\begin{eqnarray}
  \langle |\delta V (\omega)|^2 \rangle &=& \frac{1}{\omega^2} \left( \frac{1}{e^2} \frac{\partial \epsilon_F}{\partial N} \right)^2 \langle |\delta I_{J}(\omega)|^2 \rangle . \label{tot_fluct_VJ_omega}
\end{eqnarray}
\end{subequations}

Nevertheless, the NEP is not directly relevant for SPCs.
In our SPC, a temperature pulse with a time evolution similar to the ones presented in Fig.~\ref{time_evol_Te} should be discerned from the background noise and for this we need different, more specific criteria than in the integrating detectors.
In Appendix~\ref{subsec_SPC} we present a short list of such criteria, used by different authors, but here we employ the one used by us in Ref.~\cite{ApplPhysLett.82.293.2003.Anghel}, which -- we consider -- is the most appropriate.

\begin{figure}[t]
  \centering
  \includegraphics[width=9cm,keepaspectratio=true]{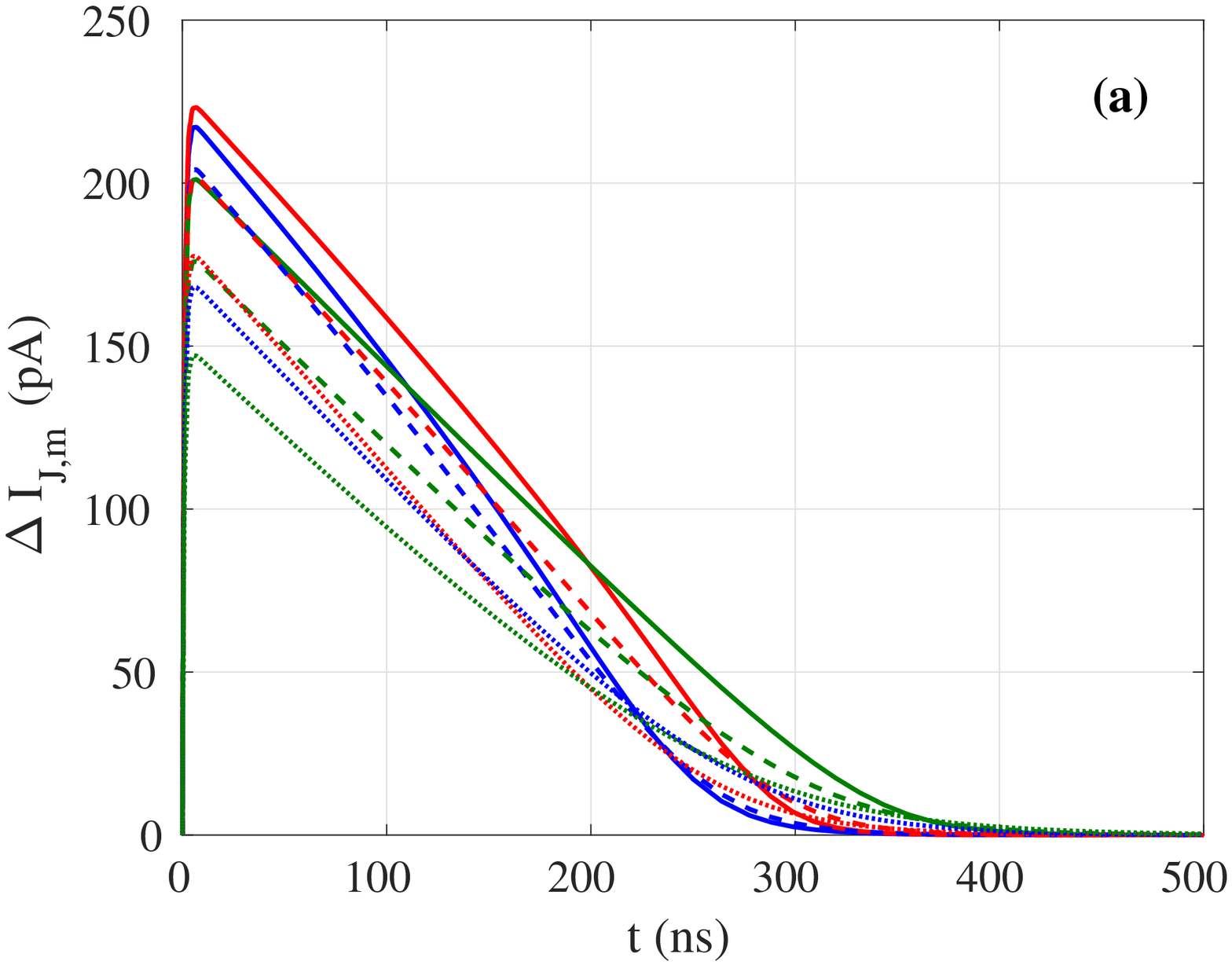}
  \includegraphics[width=9cm,keepaspectratio=true]{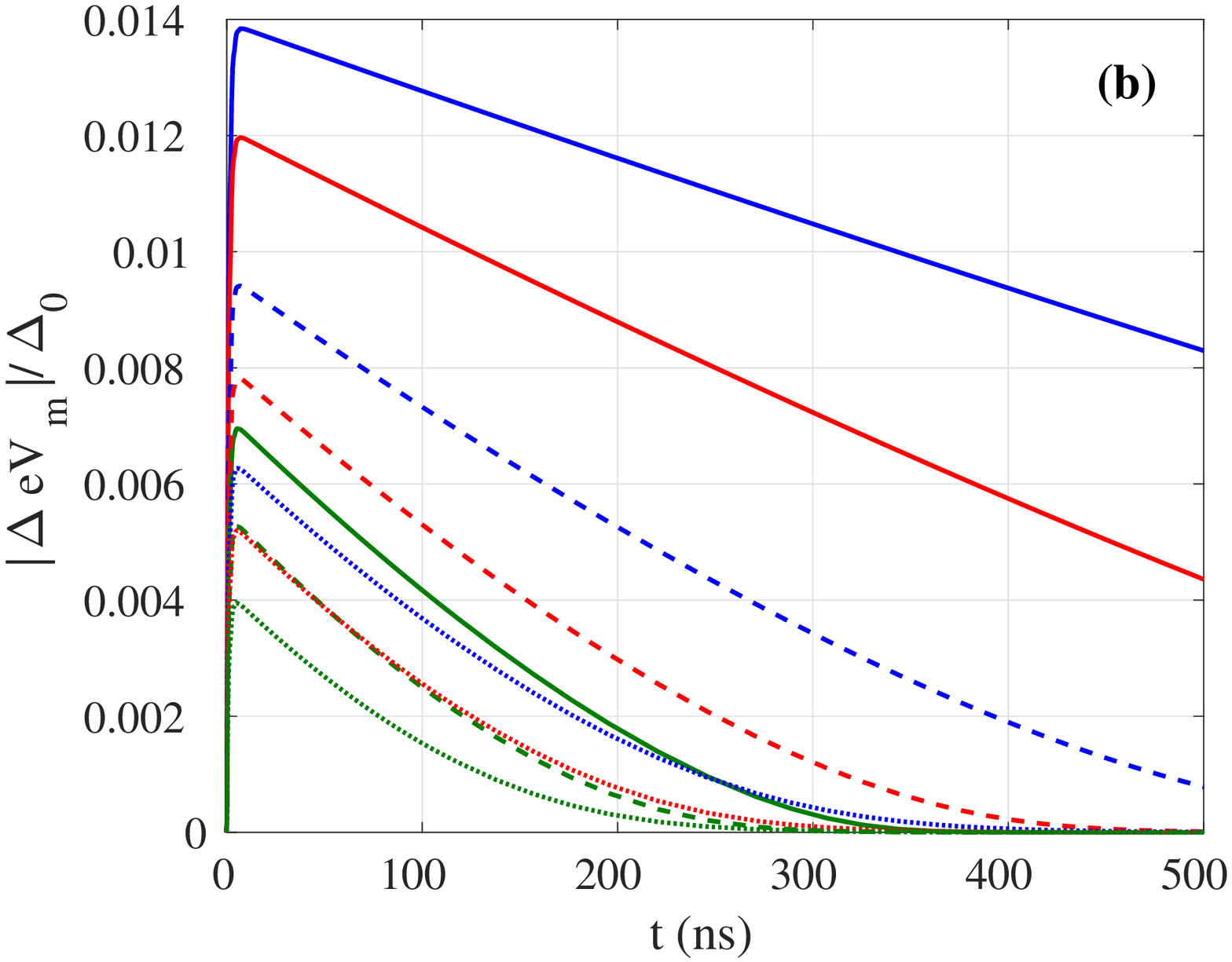}
  \caption{The time evolution of the current, at voltage-bias (a), and of the voltage, at current-bias (b), for a few significant bath temperatures and voltages (the same as in Fig.~\ref{time_evol_Te}).
  The  bath temperature takes three values: 30~mK (solid lines), 40~mK (dashed lines), and 50~mK (dotted lines).
  In (a), the voltages are fixed at $V_0$, with $1-eV_0/\Delta_0 = 0.43\times 10^{-3}$ (blue lines), $0.22\times 10^{-3}$ (red lines), and 0 (green lines).
  In (b) the currents are fixed $I_J(T_{e1}, V_0)$, where $V_0$ correspond to the same values as in (a).
  The measurement cutoff time used is $\tau_c = 1$~ns.
  }
  \label{time_evol_IJ_eV}
\end{figure}

From Eqs.~(\ref{tot_fluct_IJ_VJ_omega}) we may calculate the total fluctuation of current and voltage as
\begin{subequations} \label{delta_IJ_V_int}
\begin{eqnarray}
  \langle \delta^2 I_{J} \rangle_{tot} &=& \int_0^\infty \langle | \delta I_{J} (\omega) |^2 \rangle d\omega , \label{delta_IJ_int} \\
  \langle \delta^2 V \rangle_{tot} &=& \int_0^\infty \langle | \delta V (\omega) |^2 \rangle d\omega , \label{delta_V_int}
\end{eqnarray}
\end{subequations}
but the integrals in~(\ref{delta_IJ_V_int}) are divergent (the shot noise, for example, is ``white'', i.e. $| \delta I_{J,{\rm shot}} (\omega) |^2$ is constant) and therefore they do not represent the observed quantities.
Physically, the noise may be filtered in a band $[\omega_{max}, \omega_{min}]$, whereas the readout system may have a delay $\tau_c$, which introduces a cutoff.
Due to  the cutoff $\tau_c$, the measured quantity $M_m(t)$--which is a function of $t$--follows the real quantity $M(t)$ with a delay, modeled by the equation~\cite{ApplPhysLett.82.293.2003.Anghel}
\begin{equation}
  \frac{dM_m(t)}{dt} = \frac{M(t) - M_m(t)}{\tau_c} . \label{def_tau_c}
\end{equation}
In Fig.~\ref{time_evol_IJ_eV} we plot the time dependence of the current (at voltage-bias) and voltage (at current-bias), for the same cases as in Fig.~\ref{time_evol_Te}, without taking into account the noise.
We observe that the measured current or voltage first increases abruptly (with the time constant $\tau_c$) and then decreases as the temperature of the normal metal decreases to the equilibrium value.
Since the relaxation time is of the order of tens of nanoseconds, in the following we shall consider $\tau_c = 1$~ns (of the same order as the diffusion time) in all the numerical calculations.
We observe that the signal (the maximum value of the measured quantity) is slightly smaller than the actual jump in current or voltage produced by the photon absorption (at $T_{e2}$) and depends on $\tau_c$--the bigger $\tau_c$, the smaller the signal.

The filtering band, together to the delay time $\tau_c$, leads to the total \textit{measured} fluctuations of $I_J$ and $V$,
\begin{subequations} \label{delta_IJ_V_m}
\begin{eqnarray}
  \langle \delta^2 I_{J} \rangle_m &=& 2 \int_{\omega_{min}}^{\omega_{max}} \frac{\langle | \delta I_{J} (\omega) |^2 \rangle}{1 + \omega^2 \tau_c^2} d\omega , \label{delta_IJ_m} \\
  \langle \delta^2 V \rangle_m &=& 2 \int_{\omega_{min}}^{\omega_{max}} \frac{\langle | \delta V (\omega) |^2 \rangle}{1 + \omega^2 \tau_c^2} d\omega
  . \label{delta_V_m}
\end{eqnarray}
\end{subequations}
The relaxation time $\tau$ and the delay time $\tau_c$ impose ``filtering'' limits, even in the absence of  $\omega_{min}$ and $\omega_{max}$.
That is, the noise of frequencies much lower than the inverse of the relaxation time $\tau$ does not influence the measurement because it does not influence the visibility of the pulse in the measured quantity.
Similarly, we observe that due to $\tau_c$, the integrals~(\ref{delta_IJ_V_m}) are convergent even in the absence of $\omega_{max}$.

Plugging the expression~(\ref{fluct_I_fin}) into Eqs.~(\ref{delta_IJ_V_m}) we obtain
\begin{subequations} \label{delta_IJ_V_m2}
\begin{eqnarray}
  && \langle \delta^2 I_{J} \rangle_m
  \approx \frac{\pi}{\tau + \tau_c } \left( \frac{\partial I_J / \partial T_e}{ \partial \dot Q_T / \partial T_e} \right)^2 \left[ \langle |\delta \dot{Q}_{ep\,{\rm shot}}(\omega)|^2 \rangle \right. \nonumber \\
  && + 2 \langle |\delta\dot{Q}_{J {\rm shot}}(\omega)|^2\rangle
  \left. - \frac{4 \langle \delta I_{J {\rm shot}}(\omega) \delta \dot{Q}^*_{J {\rm shot}}(\omega) \rangle }{ S_I(0,V) } \right] \label{delta_IJ_m2} \\
  && + \left[ \frac{\pi}{\tau_c} + \frac{2}{e^2 \omega_{min}} \left(\frac{\partial\epsilon_F}{\partial N} \frac{\partial I_J}{\partial (eV)}\right)^2 \right] \langle |\delta I_{J {\rm shot}}(\omega)|^2 \rangle
  , \nonumber
\end{eqnarray}
\begin{eqnarray}
  && \langle \delta^2 V \rangle_m
  \approx \left( \frac{1}{e^2} \frac{\partial \epsilon_F}{\partial N} \right)^2 \frac{2}{\omega_{min}}
  \left\{ \left[ \langle |\delta \dot{Q}_{ep\,{\rm shot}}(\omega)|^2 \rangle
  \right. \right. \nonumber \\
  && \left. + 2 \langle |\delta\dot{Q}_{J {\rm shot}}(\omega)|^2\rangle - \frac{4 \langle \delta I_{J {\rm shot}}(\omega) \delta \dot{Q}^*_{J {\rm shot}}(\omega) \rangle }{ S_I(0,V) } \right] \nonumber \\
  && \times \left( \frac{\partial I_J / \partial T_e}{ \partial \dot Q_T / \partial T_e} \right)^2 + \langle |\delta I_{J {\rm shot}}(\omega)|^2\rangle \nonumber \\
  && \left. \left[ 1 + \frac{2}{e^2} \left(\frac{\partial\epsilon_F}{\partial N} \frac{\partial I_J}{\partial (eV)}\right)^2 \frac{1}{3 \omega_{min}^2} \right]
  \right\}
  . \label{delta_V_m2}
\end{eqnarray}
\end{subequations}
where we assumed $\tau \omega_{max}, \tau_c \omega_{max} \gg 1$ (such that $\arctan(\tau_c \omega_{max}) \approx \arctan(\tau \omega_{max}) \approx \pi/2$) and $\tau \omega_{min}, \tau_c \omega_{min} \ll 1$ (such that $\arctan(\tau_c \omega_{min}) \approx \arctan(\tau \omega_{min}) \approx 0$).

The terms that appear in Eqs.~(\ref{delta_IJ_V_m2}) are calculated in the Appendix~\ref{subsec_int_det} and in the subsection~\ref{subsubsec_NEP_lowT} we  give analytical expressions in terms of polylogarithmic functions.
In the numerical calculations we shall choose $\omega_{min} = 2\pi/(10\tau)$ (i.e., we consider that noise of frequencies lower than $1/(10\tau)$ does not obstruct the observation of the signal), where $\tau$ is plotted in Fig.~\ref{relaxation_time}.
In this way, we observe that in Eq.~(\ref{delta_IJ_m2}) the dominant terms are the ones proportional to $\langle |\delta \dot{Q}_{ep\,{\rm shot}}(\omega)|^2$ and $\langle |\delta I_{J {\rm shot}}(\omega)|^2 \rangle$, the other two terms (proportional to $\langle |\delta\dot{Q}_{J {\rm shot}}(\omega)|^2\rangle$ and $\langle \delta I_{J {\rm shot}}(\omega) \delta \dot{Q}^*_{J {\rm shot}}(\omega) \rangle$) being smaller by at least two, three orders of magnitude.
Furthermore, in the second square bracket of Eq.~(\ref{delta_IJ_m2}), the first term ($\pi/\tau_c$) is more than two orders of magnitude bigger than the second term, for the most part of the parameters range.

In the expression~(\ref{delta_V_m2}), the dominant terms are also the ones proportional to $\langle |\delta \dot{Q}_{ep\,{\rm shot}}(\omega)|^2$ and $\langle |\delta I_{J {\rm shot}}(\omega)|^2 \rangle$, with electron-phonon noise being the larger one in the most part -- not everywhere -- of the parameters range.
The sum of these two terms is two orders of magnitude larger than the sum of the terms proportional to $\langle |\delta\dot{Q}_{J {\rm shot}}(\omega)|^2\rangle$ and $\langle \delta I_{J {\rm shot}}(\omega) \delta \dot{Q}^*_{J {\rm shot}}(\omega) \rangle$.

\begin{figure}[t]
  \centering
  \includegraphics[width=9cm,keepaspectratio=true]{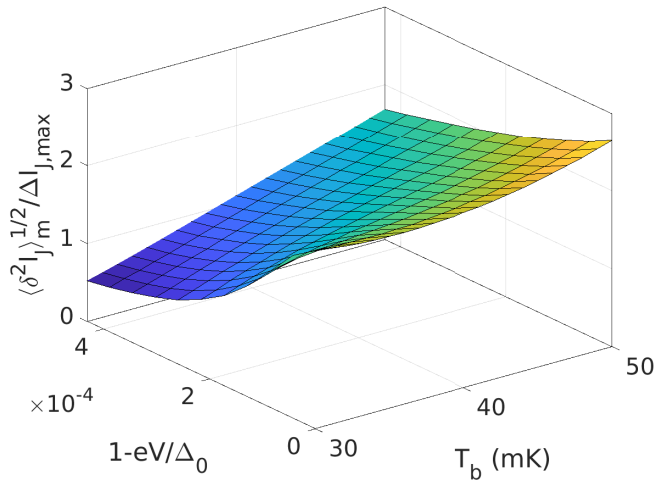}
  \caption{The current fluctuation divided by the height of the current pulse created by the photon absorption.
  We assume that the detector works as an SPC if this ratio is smaller than 1.}
  \label{rad_deltaIJ_DeltaIJ}
\end{figure}

In Fig.~\ref{rad_deltaIJ_DeltaIJ} we plot the current noise, divided by the height of the current pulse $\Delta I_{J, {\rm max}}$, as exemplified in Fig.~\ref{time_evol_IJ_eV}~a.
This ratio should be below 1, to ensure that the pulse may be observed from the noise.
We see that the SPC could function at voltage-bias in the low temperature limit (around 30~mK), but in general, in the parameters range investigated, the noise may hide the signal produced by the photon.

\begin{figure}[t]
  \centering
  \includegraphics[width=9cm,keepaspectratio=true]{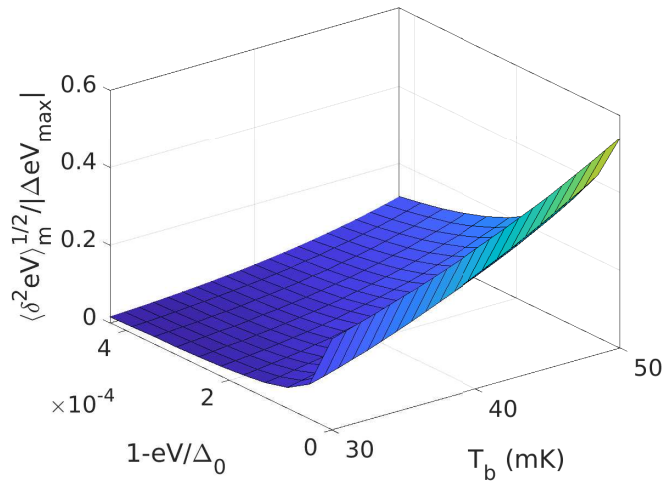}
  \caption{The voltage fluctuation divided by the height of the voltage pulse produced by the photon absorption.
  We assume that the detector works as an SPC if this ratio is smaller  than 1.}
  \label{rad_deltaV_DeltaV}
\end{figure}

The situation is better for current-biased measurements.
In Fig.~\ref{rad_deltaV_DeltaV} we plot the voltage noise divided by the height of the voltage pulse (see Fig.~\ref{time_evol_IJ_eV}~b).
We see that the detector permits the counting of 1~cm wavelength photons in the whole range of parameters analyzed.

\section{Conclusions} \label{sec_concl}

We investigated the possibility of using the cold-electron bolometer (CEB) as counter for photons of wavelengths up to 1~cm.
The CEB consists of a normal metal island, coupled to the superconducting antenna by two symmetric normal metal-insulator-superconductor (NIS) tunnel junctions, realizing the so called SINIS structure (see Fig.~\ref{SINIS_scheme}).
We presented the general formalism and the numerical calculations, for a bath temperature ($T_b$) range from 30~mK to 50~mK.
In this temperature range, the flux of 1~cm wavelength photons which hit the detector, coming from the environment, is less than 1 photon in 100 hours (see Fig.~\ref{photons_flux}).
This makes the detector suitable for counting low energy photons generated by rare events.

We investigated both, the voltage-biased setup--when the measured quantity is the current--and the current-biased setup--when the measured quantity is the voltage.
We compared the response of the detector (shown in Figs.~\ref{time_evol_Te} and \ref{time_evol_IJ_eV}) with the fluctuation of the measured quantity (see Figs.~\ref{rad_deltaIJ_DeltaIJ}-\ref{rad_deltaV_DeltaV_0o1}).

Due to the intrinsic shot noise of the current, the voltage-bias setup is more noisy than the current-bias setup.
For a volume $\Omega = 0.01~{\rm \mu m}^3$, both, current-biased and voltage-biased CEB's may detect 1~cm wavelength photons, but the current-biased detectors are more accurate.
The signal (due to photon absorption) is bigger than the noise in the current-biased CEB for the whole temperature range investigated (see Fig.~\ref{rad_deltaV_DeltaV}) whereas for the voltage-biased CEB's, the signal is bigger than the noise only in a part of the parameters range, as can be seen in Fig.~\ref{rad_deltaIJ_DeltaIJ}.

If the volume of the normal metal island is $\Omega = 0.1~{\rm \mu m}^3$, in the temperature range investigated, the noise in current in the voltage-biased setup is bigger than the current pulse produced by the photon absorption (see Fig.~\ref{rad_deltaIJ_DeltaIJ_0o1}).
Therefore, the detection of 1~cm wavelength photons seems not to be possible.
On the other hand, in the current-biased setup, the counter may work,  since the voltage pulse produced by the photon absorption is bigger than the noise, at least in some ranges of the parameters, as seen in Fig.~\ref{rad_deltaV_DeltaV_0o1}.
Nevertheless, the system may be improved.
For example, by reducing the tunneling resistance $R_T$, the cooling properties of the NIS junctions improve (if  the quasiparticles in the superconductor are  properly removed from the vicinity of the junction) and the detector becomes more sensitive.
On the other hand, the reduction of the tunneling resistance would increase the noise as well, so the optimization of the SPC is not straightforward and require a detailed analysis.

\section*{Acknowledgements}

The authors would like to thank Giovanni Carugno and Claudio Gatti for stimulating discussions on the development of a single photon counter for searching axions.
The work was supported by RSF (project 16-19-10468, the main conceptual ideas, contribution to the design of samples, and interpretation of the results by LK) and by the ANCS (project PN19060101/2019, modeling, analytical and numerical calculations, and interpretation of results, by DVA) and Romania-JINR collaboration projects.

\appendix

\section{Results for a volume of the normal metal of 0.1~${\rm \mu m}^3$}

\begin{figure}[t]
  \centering
  \includegraphics[width=9cm,keepaspectratio=true]{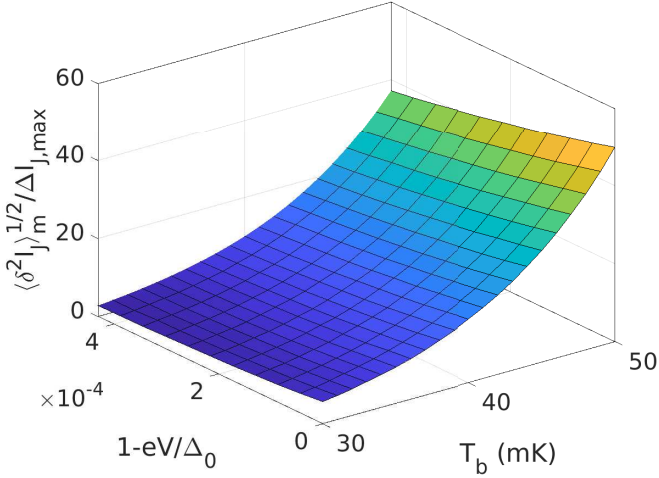}
  \caption{The current fluctuation divided by the height of the current pulse created by the photon absorption, for $\Omega = 0.1~{\rm \mu m}^3$.
  The fluctuation is bigger than the pulse for the whole range of parameters.}
  \label{rad_deltaIJ_DeltaIJ_0o1}
\end{figure}

\begin{figure}[t]
  \centering
  \includegraphics[width=9cm,keepaspectratio=true]{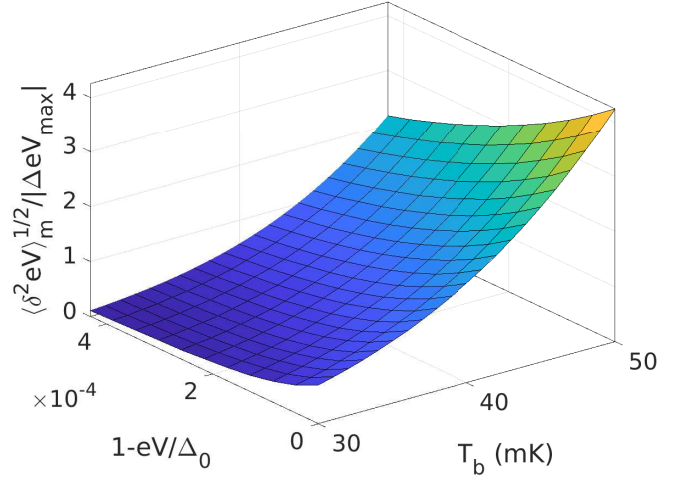}
  \caption{The voltage fluctuation divided by the height of the (absolute value of the) voltage pulse created by the photon absorption, for $\Omega = 0.1~{\rm \mu m}^3$.
  The fluctuation is smaller than the pulse -- so the detector may work as an SPC -- in the lower part of the temperature range.
  }
  \label{rad_deltaV_DeltaV_0o1}
\end{figure}

For comparison, we present the relative fluctuations of current and voltage for a detector with $\Omega = 0.1~{\rm \mu m}^3$.
The  relative fluctuation of current (the fluctuation of current, divided by the current pulse) in the voltage-bias setup is presented in Fig.~\ref{rad_deltaIJ_DeltaIJ_0o1} and the relative fluctuation of voltage (voltage fluctuation divided by the voltage pulse) is presented in Fig.~\ref{rad_deltaV_DeltaV_0o1}.
We observe that in the voltage-bias setup $\sqrt{\langle \delta^2 I_J\rangle_m}/\Delta I_{J, {\rm max}} > 1$, so the signal is smaller than the noise.
In principle, in such a situation the signal cannot be distinguished from the noise.
On the other hand, in the current-bias setup, in some ranges of parameters, the signal is bigger than the noise (i.e. $\sqrt{\langle \delta^2 (eV)\rangle_m}/|\Delta eV_{\rm max}| < 1$) and the device may function as a photon counter, even for such relatively large volume.

\section{Fluctuations and the NEP in integrating detectors and SPCs} \label{sec_current_fluct}

\subsection{Integrating detectors} \label{subsec_int_det}

Noise in integrating detectors have been calculated before (see, for example, Refs.~\cite{JLowTempPhys.123.197.2001.Anghel, JApplPhys.89.6464.2001.Golubev}).
The figure of merit for such detectors is the NEP, which represents the signal power which gives the signal to noise ratio equal to 1.
To calculate the NEP, one has to calculate the added contributions of all the sources of noise in the detector.
In Ref.~\cite{JApplPhys.89.6464.2001.Golubev}, the NEP of a voltage biased junction was given as
(we slightly change the notations, to adapt them to ours)
\begin{subequations} \label{def_NEP_GK}
\begin{equation}
  {\rm NEP}^2 \equiv
  {\rm NEP}^2_{ep} + {\rm NEP}^2_{J} ,
  \label{NEP_GK_gen}
\end{equation}
where we omitted a term $\langle |\delta I_{J {\rm amp}} |^2 \rangle/S_I^2 (0,V)$ ($S_I$ is defined in Eq.~\ref{I_resp}), representing the contribution due to the sensitivity of the amplifier, which we cannot calculate and therefore we do not take into account (as mentioned before) -- this term has to be added to the final result.
In Eq.~(\ref{NEP_GK_gen}), ${\rm NEP}^2_{J}$ is the noise associated to the transport through the junctions (in Ref.~\cite{JApplPhys.89.6464.2001.Golubev} there was only one NIS junction, whereas we have two) and ${\rm NEP}^2_{ep}$ is the  shot noise due to electron-phonon interaction.
In our model, both junctions are identical and, according to~\cite{JApplPhys.89.6464.2001.Golubev}, each contributes with
\begin{eqnarray}
  \frac{{\rm NEP}^2_{J}}{2} &=& \langle |\delta \dot Q_{J {\rm shot}}(\omega)|^2 \rangle - 2 \frac{\langle \delta \dot Q_{J {\rm shot}}(\omega) \delta I^*_{J {\rm shot}}(\omega) \rangle}{S_I (0,V)} \nonumber \\
  && + \frac{\langle |\delta I_{J {\rm shot}}(\omega)|^2 \rangle}{S_I^2 (0,V)}
  \label{NEP_GK_J1}
\end{eqnarray}
whereas~\cite{ProcLTD7.64.1997.Golwala, JApplPhys.89.6464.2001.Golubev}
\begin{equation}
  {\rm NEP}^2_{ep} = 10 k_B \Sigma_{ep} \Omega (T_e^6 + T_{ph}^6) \equiv \langle |\delta \dot{Q}_{ep}(\omega)|^2 \rangle
  \label{NEP_GK_ep}
\end{equation}
(in our  case, $T_{ph} = T_b$).
The terms appearing in Eq.~(\ref{NEP_GK_J1}) are the shot noise of the heat power through one NIS junctions~\cite{JLowTempPhys.123.197.2001.Anghel, JApplPhys.89.6464.2001.Golubev},
\begin{eqnarray}
&& \langle |\delta\dot{Q}_{J {\rm shot}} (\omega)|^2 \rangle = 2
\int_\Delta^\infty [(\epsilon-eV)^2(j_1+j_3) - (\epsilon+eV)^2 \nonumber \\
&& \times (j_4+j_2)]\, d\epsilon
, \label{pshot_1}
\end{eqnarray}
the current shot noise,
\begin{equation}
  \langle |\delta I_{J {\rm shot}(\omega)}|^2 \rangle = 2 e^2 \int_\Delta^\infty [j_1(\epsilon) + j_2(\epsilon) + j_3(\epsilon) + j_4(\epsilon)] \label{current_gen}
\end{equation}
the correlation between the  heat power shot noise and the current shot noise,
\begin{eqnarray}
&& \langle \delta\dot{Q}_{J {\rm shot}} (\omega) \delta I^*_{J {\rm shot}}(\omega) \rangle = 2 e
\int_\Delta^\infty [(\epsilon-eV)(j_1+j_3) \nonumber \\
&& - (\epsilon+eV) (j_4+j_2)]\, d\epsilon
= \langle \delta\dot{Q}^*_{J {\rm shot}} (\omega) \delta I_{J {\rm shot}}(\omega) \rangle,
\label{pI_shot}
\end{eqnarray}
and the current responsivity,
\begin{equation}
  S_I(\omega, V) = \frac{\partial I_J / \partial T_e}{ - i \omega C_V + \partial \dot Q_T / \partial T_e}
  \equiv 
   \frac{S_I(0,V)}{ - i \omega \tau + 1 } ,
  \label{I_resp}
\end{equation}
\end{subequations}
where $S_I(0, V) \equiv (\partial I_J / \partial T_e) / (\partial \dot Q_T / \partial T_e)$.

Nevertheless, in Ref.~\cite{JLowTempPhys.123.197.2001.Anghel} it was shown that other terms contribute to the total fluctuations, as we explain below.
The fluctuation of the power exchange between the electrons of the normal metal and the other subsystems lead to fluctuations of the electrons temperature, whereas the fluctuation of (electrons) current leads to the fluctuation of the Fermi energy, which may be interpreted as a fluctuation of the applied voltage.
These fluctuations should also be taken into account in the calculation of the NEP of the detector, since they lead to extra contributions to the fluctuations of the measured quantity $I_J$.
In this way, one arrives to a self-consistent set of equations, which may be solved for an exact solution, or one may only plug in the different sources of shot noise, for a perturbative calculation.
These extra contributions may be obtained by writing the detailed energy balance equations in the normal metal (see Ref.~\cite{JLowTempPhys.123.197.2001.Anghel} for details).
Applying the Fourier transformation, these equations may be formally written as
\begin{subequations} \label{def_Deltas}
\begin{equation}
  \Delta_T (\omega) = \Delta_T'(\omega) \label{Delta_Deltap}
\end{equation}
where the fluctuation of temperature
\begin{eqnarray}
  \Delta_T(\omega) &=& \delta T_e(\omega) \left[ - i\omega C_V + \frac{\partial}{\partial T_e} (2 \dot{Q}_J + \dot{Q}_{ep}) \right] \nonumber \\
  &\equiv& \delta T_{e}(\omega) Z_T \label{delta1}
\end{eqnarray}
is related to the fluctuation of power
\begin{eqnarray}
  && \Delta'_T(\omega) = - \delta \dot{Q}_{J1}(\omega) - \delta \dot{Q}_{J2}(\omega) - \delta \dot{Q}_{ep}(\omega) + \delta \dot{Q}_{oe}(\omega) \nonumber \\
  & & + \frac{1}{e} \frac{\partial\epsilon_{\rm F}}{\partial N} \frac{\partial \dot{Q}_{J1}}{\partial (eV)}\frac{\delta I_{J1}(\omega)}{i\omega}
  + \frac{1}{e} \frac{\partial\epsilon_{\rm F}}{\partial N} \frac{\partial \dot{Q}_{J2}}{\partial (eV)}\frac{\delta I_{J2}(\omega)}{i\omega} \label{delta2}
\end{eqnarray}
\end{subequations}
and we defined the thermal impedance of the electron system, $Z_T(\omega)$ -- in Eq.~(\ref{delta2}) we took into account that we have two junctions, $J1$ and $J2$.
Comparing Eqs.~(\ref{delta1}) and (\ref{I_resp}), we observe that
\begin{equation}
  Z_T(\omega) = \frac{1}{S_I(\omega, V)} \frac{\partial I_J}{\partial T_e} .
  \label{rel_ZT_SI}
\end{equation}
In addition to Ref.~\cite{JApplPhys.89.6464.2001.Golubev}, one should also take into account the fluctuation due to the incoming radiation, denoted here by $\dot Q_{oe}(\omega)$, since this produces (at least) shot noise.
Equations~(\ref{def_Deltas}) are a simplified version of Eqs.~(10)-(12) of Ref.~\cite{JLowTempPhys.123.197.2001.Anghel}.
The difference comes from the fact that here we consider $T_{ph} = T_b =$constant and we assume that the incoming radiation does not directly interact with the lattice (phonons).
Under these simplifying assumptions, calculating the mean square fluctuations gives
\begin{subequations} \label{deltas_2}
\begin{eqnarray}
  && \langle |\Delta_T(\omega)|^2 \rangle = \langle\delta^2 T_e\rangle_\omega \left| Z_T(\omega) \right|^2  \nonumber \\ 
  && = \langle |\delta \dot{Q}_{ep\,{\rm shot}}(\omega)|^2 \rangle
  + \langle |\delta \dot{Q}_{oe\,{\rm shot}}(\omega)|^2 \rangle + \Upsilon (\omega)
  \nonumber \\
  && \equiv \langle |\Delta'_T(\omega)|^2 \rangle , \label{delta2p}
\end{eqnarray}
where $\delta \dot{Q}_{ep\,{\rm shot}}(\omega)$ is the shot noise of the incoming (optical) power and we denoted
\begin{eqnarray}
  \Upsilon(\omega) &\equiv& 2 \left\langle \left| - \delta \dot{Q}_{J\,{\rm shot}}(\omega)
  + \frac{1}{e} \frac{\partial\epsilon_{\rm F}}{\partial N}\frac{\partial \dot{Q}_J}{\partial (eV)} \frac{\delta I_{J\,{\rm shot}}(\omega)}{i\omega} \right|^2 \right\rangle \nonumber \\
  &=& 2 \langle |\delta\dot{Q}_{J {\rm shot}}(\omega)|^2\rangle + \frac{2}{\omega^2 e^2}\left(\frac{\partial\epsilon_F}{\partial N} \frac{\partial \dot{Q}_{\rm J}}{\partial (eV)}\right)^2 \nonumber \\
  && \times \langle |\delta I_{J {\rm shot}}(\omega)|^2\rangle .
  \label{upsi}
\end{eqnarray}
\end{subequations}
In Eq.~(\ref{upsi}) we took into account that the correlation between $\delta \dot{Q}_{J {\rm shot}} (\omega)$ and  $\delta I_{J {\rm shot}} (\omega) / i\omega$ is zero because of the $\pi/2$ phase difference.
In Eq.~(\ref{upsi}) we dropped the separate subscripts $J1$ and $J2$ used in Eq.~(\ref{delta2}), by making the simplifying assumption that the noise contributions are uncorrelated between the two junctions.

From Eqs.~(\ref{def_Deltas})-(\ref{deltas_2}) we can calculate the spectral density of the temperature noise, which may be used further, for the calculation of the fluctuation of the junction current, as shown in Eq.~(\ref{tot_fluct_IJ_omega}).
In Eq.~(\ref{tot_fluct_IJ_omega}) we used $\Re \left[\left\langle |\delta I_J(\omega)|^2 / (i e\omega) \right\rangle \right] = 0$ and we assumed that the changes in the chemical potential $\mu$ due to temperature fluctuations are negligible.
The correlations $\Re \left(\langle \delta I_J(\omega) \delta T^*_e(\omega) \rangle \right)$ and $\Re \left(\left\langle \delta I_J(\omega) \delta T^*_e (\omega)/(i e\omega) \right\rangle \right) = \Im \left(\left\langle \delta I_J(\omega) \delta T^*_e (\omega) \right\rangle \right)/(e\omega)$ should be calculated using Eqs.~(\ref{def_Deltas}) -- by $ \Im (\cdot)$ we denote the imaginary part of a complex number.
Having the fluctuation of the current we can calculate the NEP.
For this, we have to calculate the energy of the incoming radiation (in the unit bandwidth), which produces a signal equal to the fluctuation of current~(\ref{tot_fluct_IJ_omega}).
The incoming radiation produces a change in the electrons temperature $\delta_s T_e(\omega)$, which produces a signal $\delta_s I_J(\omega)$.
The two quantities are related by
\begin{equation}
  \delta_s I_J(\omega) = \frac{\partial I_J}{\partial T_e} \delta_s T_e \equiv \frac{\partial I_J}{\partial T_e} \frac{\dot Q_{oe}}{Z_T(\omega)}
  \equiv S_I(\omega, V) Q_{oe} . \label{signal_amp}
\end{equation}
Equating $|\delta_s I_J(\omega)|^2$ with $\langle |\delta I_J(\omega)|^2 \rangle$ obtained from~(\ref{tot_fluct_IJ_omega}), we finally get the full expression of the NEP,
\begin{eqnarray}
  && {\rm NEP}_t^2(\omega) \equiv \frac{\langle |\delta I_J(\omega)|^2 \rangle}{|S_I(\omega,V)|^2} = \langle |\Delta'_T(\omega)|^2 \rangle
  + 2 \frac{\langle |\delta I_{J {\rm shot}}(\omega)|^2 \rangle}{|S_I(\omega,V)|^2} \nonumber \\
  && + \frac{2}{e^2 \omega^2}\left(\frac{\partial\epsilon_F}{\partial N} \frac{\partial I_J}{\partial (eV)}\right)^2 \frac{\langle |\delta I_J(\omega)|^2 \rangle}{|S_I(\omega,V)|^2}
  + \frac{2 Corr_{\rm \omega}}{|S_I(\omega,V)|^2} , \label{nept}
\end{eqnarray}
where $Corr(\omega)$ denotes the two correlation terms from Eq. (\ref{tot_fluct_IJ_omega}).
From Eq.~(\ref{nept}) we arrive to
\begin{eqnarray}
  && {\rm NEP}_t^2(\omega) = \langle |\delta \dot{Q}_{ep\,{\rm shot}}(\omega)|^2 \rangle
  + \langle |\delta \dot{Q}_{oe\,{\rm shot}}(\omega)|^2 \rangle \nonumber \\
  && + 2 \langle |\delta\dot{Q}_{J {\rm shot}}(\omega)|^2\rangle
   + 2 \frac{\langle |\delta I_{J {\rm shot}}(\omega)|^2 \rangle}{|S_I(\omega,V)|^2}\nonumber \\
  && + \frac{2}{\omega^2 e^2}\left(\frac{\partial\epsilon_F}{\partial N} \frac{\partial \dot{Q}_{\rm J}}{\partial (eV)}\right)^2 \langle |\delta I_{J {\rm shot}}(\omega)|^2\rangle \nonumber \\
  && + \frac{2}{e^2 \omega^2}\left(\frac{\partial\epsilon_F}{\partial N} \frac{\partial I_J}{\partial (eV)}\right)^2 \frac{\langle |\delta I_{J}(\omega)|^2 \rangle}{|S_I(\omega,V)|^2} \label{nept2} \\
  && + 4 \frac{\partial I_J}{\partial T_e} \frac{\Re \left(\langle \delta I_J(\omega) \delta T^*_e(\omega) \rangle \right)}{|S_I(\omega,V)|^2} \nonumber \\
  && + \frac{4}{e \omega} \frac{\partial\epsilon_{\rm F}}{\partial N} \frac{\partial I_J}{\partial (eV)} \frac{\partial I_J}{\partial T_e} \frac{\Im \left(\langle \delta I_J(\omega) \delta T^*_e (\omega) \rangle \right)}{|S_I(\omega,V)|^2} , \nonumber
\end{eqnarray}
Using Eqs.~(\ref{def_Deltas}), we calculate in the  lowest order (i.e. taking only the shot noise contributions to $\delta I_{J}(\omega) \equiv \delta I_{J {\rm shot}}(\omega)$ and $\delta \dot Q_J(\omega) \equiv \delta \dot Q_{J {\rm shot}}(\omega)$) the correlation
\begin{subequations} \label{calc_cor}
\begin{eqnarray}
  && \langle \delta I_J(\omega) \delta T^*_e (\omega) \rangle =
  \left\langle \frac{\delta I_{J {\rm shot}}(\omega)}{Z^*_T(\omega)}
  \left[ - \delta \dot{Q}_{J {\rm shot}}(\omega) \right. \right. \nonumber \\
  && \left. \left. + \frac{1}{e} \frac{\partial\epsilon_{\rm F}}{\partial N} \frac{\partial \dot{Q}_{J1}}{\partial (eV)}\frac{\delta I_{J {\rm shot}}(\omega)}{i\omega} \right]^* \right\rangle \nonumber \\
  && = - \left\langle \frac{\delta I_{J {\rm shot}}(\omega) \delta \dot{Q}^*_{J {\rm shot}}(\omega)}{Z^*_T(\omega)} \right\rangle \nonumber \\
  && - \frac{1}{e} \frac{\partial\epsilon_{\rm F}}{\partial N} \frac{\partial \dot{Q}_{J1}}{\partial (eV)} \left\langle \frac{|\delta I_{J {\rm shot}}(\omega)|^2}{i\omega Z^*_T(\omega)} \right\rangle .
  \label{calc_cor1}
\end{eqnarray}
Using (\ref{calc_cor1}), we obtain
\begin{eqnarray}
  && \Re [\langle \delta I_J(\omega) \delta T^*_e (\omega) \rangle ]
  = - \frac{Z_T(0)}{|Z_T(\omega)|^2} \langle \delta I_{J {\rm shot}}(\omega) \delta \dot{Q}^*_{J {\rm shot}}(\omega) \rangle \nonumber \\
  && + \frac{1}{e} \frac{\partial\epsilon_{\rm F}}{\partial N} \frac{\partial \dot{Q}_{J1}}{\partial (eV)} \frac{C_V}{|Z_T(\omega)|^2} \langle |\delta I_{J {\rm shot}}(\omega)|^2 \rangle .
  \label{calc_cor_re}
\end{eqnarray}
and
\begin{eqnarray}
  && \Im [\langle \delta I_J(\omega) \delta T^*_e (\omega) \rangle]
  = \frac{\omega C_V}{|Z_T(\omega)|^2} \langle \delta I_{J {\rm shot}}(\omega) \delta \dot{Q}^*_{J {\rm shot}}(\omega) \rangle \nonumber \\
  && + \frac{1}{e} \frac{\partial\epsilon_{\rm F}}{\partial N} \frac{\partial \dot{Q}_{J}}{\partial (eV)} \frac{Z_T(0)}{\omega |Z_T(\omega)|^2} \langle |\delta I_{J {\rm shot}}(\omega)|^2 \rangle
  \label{calc_cor_im}
\end{eqnarray}
\end{subequations}
(notice that $\langle \delta I_{J {\rm shot}}(\omega) \delta \dot{Q}^*_{J {\rm shot}}(\omega) \rangle$ is real, as specified in Eq.~\ref{pI_shot}).
Combining Eq.~(\ref{nept2}) with Eqs.~(\ref{calc_cor}), we obtain the expression for the total noise, which we formally write as
\begin{subequations} \label{NEP_tot}
\begin{eqnarray}
  && {\rm NEP}_t^2(\omega) = \sum_{i=0}^{8} {\rm NEP}_i \label{NEP_tot_gen}
\end{eqnarray}
where
\begin{eqnarray}
  && {\rm NEP}_0 \equiv \langle |\delta \dot{Q}_{ep\,{\rm shot}}(\omega)|^2 \rangle
  + \langle |\delta \dot{Q}_{oe\,{\rm shot}}(\omega)|^2 \rangle \label{NEP0} \\
  && {\rm NEP}_1 \equiv 2 \langle |\delta\dot{Q}_{J {\rm shot}}(\omega)|^2\rangle \label{NEP1} \\
  && {\rm NEP}_2 \equiv 2 \frac{\langle |\delta I_{J {\rm shot}}(\omega)|^2 \rangle}{|S_I(\omega,V)|^2}\label{NEP2} \\
  && {\rm NEP}_3 \equiv - 4 \langle \delta I_{J {\rm shot}}(\omega) \delta \dot{Q}^*_{J {\rm shot}}(\omega) \rangle / S_I(0,V) \label{NEP3} \\
  && {\rm NEP}_4 \equiv \frac{2}{e^2 \omega^2}\left(\frac{\partial\epsilon_F}{\partial N} \frac{\partial \dot{Q}_{\rm J}}{\partial (eV)}\right)^2 \langle |\delta I_{J {\rm shot}}(\omega)|^2\rangle \label{NEP4} \\
  && {\rm NEP}_5 \equiv \frac{2}{e^2 \omega^2}\left(\frac{\partial\epsilon_F}{\partial N} \frac{\partial I_J}{\partial (eV)}\right)^2 \frac{\langle |\delta I_{J {\rm shot}}(\omega)|^2 \rangle}{|S_I(\omega,V)|^2} \label{NEP5} \\
  && {\rm NEP}_6 \equiv \frac{4 C_V}{e} \left( \frac{\partial I_J}{\partial T_e} \right)^{-1} \frac{\partial\epsilon_{\rm F}}{\partial N} \frac{\partial \dot{Q}_{J}}{\partial (eV)} \langle |\delta I_{J {\rm shot}}(\omega)|^2 \rangle
  \label{NEP6} \\
  && {\rm NEP}_7 \equiv \frac{4 C_V}{e} \frac{\partial\epsilon_{\rm F}}{\partial N} \frac{\partial I_J}{\partial (eV)} \left( \frac{\partial I_J}{\partial T_e} \right)^{-1} \nonumber \\
  && \times \langle \delta I_{J {\rm shot}}(\omega) \delta \dot{Q}^*_{J {\rm shot}}(\omega) \rangle \label{NEP7} \\
  && {\rm NEP}_8 \equiv \frac{4}{e^2 \omega^2} \left( \frac{\partial\epsilon_{\rm F}}{\partial N} \right)^2 \frac{\partial I_J}{\partial (eV)} \frac{\partial \dot{Q}_{J}}{\partial (eV)} \frac{\langle |\delta I_{J {\rm shot}}(\omega)|^2 \rangle}{S_I(0, V)}
  , \label{NEP8}
\end{eqnarray}
\end{subequations}
We notice that the terms~(\ref{NEP0})-(\ref{NEP3}) contain all the terms of Eqs.~(\ref{def_NEP_GK})~\cite{JApplPhys.89.6464.2001.Golubev}.
But, beside the term $\langle |\delta \dot{Q}_{oe\,{\rm shot}}(\omega)|^2 \rangle$, which was not taken into account in~(\ref{def_NEP_GK}), we notice another difference.
The term ${\rm NEP}_2 \equiv 2 \langle |\delta I_{J {\rm shot}}(\omega)|^2 \rangle / |S_I(\omega,V)|^2$, which \textit{is} $\omega$ \textit{dependent}, appears in Eqs.~(\ref{def_NEP_GK}) -- and in Ref.~\cite{JApplPhys.89.6464.2001.Golubev} -- as ${\rm NEP}_2' \equiv 2 \langle |\delta I_{J {\rm shot}}(\omega)|^2 \rangle / |S_I(0,V)|^2$, which \textit{is} $\omega$ \textit{independent}.
This mathematical difference, although well justified in our formalism, has also a simple intuitive explanation: to produce the same current oscillations, the amplitude of the power oscillations should increase with frequency due to the thermal inertia of the electron gas.
This leads to the increase of the contribution of $\delta I_{J {\rm shot}}(\omega)$ with $\omega$ in our expression for NEP.

\subsubsection{Evaluation of the contributions to the NEP in the low temperature limit} \label{subsubsec_NEP_lowT}

Let us derive simplified expressions for the terms of ${\rm NEP}_t^2(\omega)$~(\ref{NEP_tot}), valid at low temperatures, $T_e, T_b \ll \Delta/k_B \, (\approx 2.32$~K, for $\Delta = 0.2$~meV in Al).
At such temperatures we ignore the currents of holes and the population of quasiparticle states in the superconductor by using the approximation
\begin{equation}
  f (\epsilon + eV, T_e) \approx f(\epsilon, T_b) \approx 0 .
  \label{approx_fTb}
\end{equation}
Then, the current $I_J$ through one junction~(\ref{def_IJ}) may be simplified to
\begin{equation}
  I_J \approx \frac{1}{e R_T} \int_\Delta^\infty \frac{\epsilon}{\sqrt{\epsilon^2-\Delta^2}} \frac{d \epsilon}{e^{\beta_e(\epsilon - eV)} + 1} . \label{current_J}
\end{equation}
We introduce  the notations $x \equiv \beta_e (\epsilon - \Delta)$, $a_e \equiv \beta_e (\Delta - eV)$, and $A_e \equiv \beta_e \Delta$, to write
\begin{eqnarray}
  I_J &=& \frac{1}{e R_T} \frac{k_B T}{\sqrt{2 A_e}} \int_0^\infty \frac{x + A_e}{\sqrt{x[ 1 + x/(2A_e)]}} \frac{d x}{e^{x + a_e} + 1} . \nonumber
\end{eqnarray}
Using the Taylor expansion $1/\sqrt{1+\delta} \approx 1 - \delta/2$ (in this case, $\delta \equiv x/(2A_e)$), we obtain
\begin{eqnarray}
  I_J &\approx& \frac{1}{e R_T} \frac{(k_B T)^{3/2}}{\sqrt{2 \Delta}} \int_0^\infty \left( \sqrt{x} + \frac{A_e}{\sqrt{x}} \right) \left( 1 - \frac{x}{4A_e} \right) \nonumber \\
  && \times \frac{d x}{e^{x + a_e} + 1}
  \approx \frac{1}{e R_T} \frac{(k_B T)^{3/2}}{\sqrt{2 \Delta}} \left[ \frac{3}{4} \int_0^\infty \frac{\sqrt{x} \, d x}{e^{x + a_e} + 1} \right. \nonumber \\
  && \left. + A_e \int_0^\infty \frac{dx}{\sqrt{x} \left( e^{x+a_e} + 1 \right)} \right]
  \equiv - \frac{\sqrt{k_B T \Delta}}{e R_T} \sqrt{\frac{\pi}{2}} \nonumber \\
  && \times \left[ Li_{\frac{1}{2}}\left( - e^{-a_e} \right) + \frac{3}{8} \frac{k_B T}{\Delta} Li_{\frac{3}{2}}\left( - e^{-a_e} \right) \right] , \label{current_J2}
\end{eqnarray}
where we kept only the two highest order terms in $A_e$ and $Li_m(x)$ is the polylogarithm of order $m$ in $x$~\cite{Lewin:book, JMathPhys36.1217.1995.Lee, PhysA304.421.2002.Lee, JPA35.7255.2002.Anghel} (notice that $Li_{l}\left(-e^{-a_e} \right) < 0$, therefore the global ``$-$'' sign in the equations involving polylogarithms).
Similarly, the heat power through a NIS junction~(\ref{def_QJ}) is reduced to~\cite{JLowTempPhys.123.197.2001.Anghel}
\begin{eqnarray}
  \dot Q_J &\approx& \frac{1}{e^2 R_T} \int_\Delta^\infty \frac{(\epsilon-eV) \epsilon}{\sqrt{\epsilon^2-\Delta^2}} \frac{d \epsilon}{e^{\beta_e(\epsilon - eV)} + 1} \nonumber \\
  &\approx& - \frac{1}{e^2 R_T} \sqrt{\frac{\pi (k_B T)^3 \Delta}{2}} \left[ \frac{1}{2} \left( 1 + \frac{3}{4} \frac{a_e}{A_e} \right) Li_{3/2} (-e^{-a_e}) \right. \nonumber \\
  && \left. + a_e Li_{1/2} (-e^{-a_e}) + \frac{3}{4 A_e} Li_{5/2} (-e^{-a_e}) \right] . \label{heat_J}
\end{eqnarray}
If one neglects the terms proportional to $1/A_e$ in the square brackets of Eq.~(\ref{heat_J}), one can obtain the so called \textit{optimum cooling power}, for $a_e \approx 0.66$.

The current shot noise through one junction is given by Eq.~(\ref{current_gen}).
Ignoring again the tunneling of holes and the populations of the quasiparticle states in the superconductor, this is reduced to
%
\begin{eqnarray}
  \langle | \delta I_{J {\rm shot}} (\omega) |^2 \rangle
  &\approx& \frac{2 e^2}{e^2 R_T} \int_\Delta^\infty \frac{\epsilon}{\sqrt{\epsilon^2 - \Delta^2}} \frac{d\epsilon}{e^{\beta_e (\epsilon - eV)} + 1} \nonumber \\
  &\equiv& 2e I_J . \label{IJ_shot}
\end{eqnarray}
%
%

The heat current fluctuation through one junction, in the approximation (\ref{approx_fTb}), is
\begin{eqnarray}
  && \langle | \delta \dot Q_{J {\rm shot}} (\omega) |^2 \rangle \approx \frac{2}{e^2 R_T} \int_\Delta^\infty \frac{\epsilon (\epsilon - eV)^2}{\sqrt{\epsilon^2 - \Delta^2}} \frac{d\epsilon}{e^{\beta_e (\epsilon - eV)} + 1} \nonumber \\
  &&\approx - \frac{2 (k_B T)^{5/2}}{e^2 R_T} \sqrt{\frac{\pi \Delta}{2}} \left\{ \frac{45}{32 A_e} Li_{7/2} \left( - e^{-a_e} \right) \right. \nonumber \\
  && + \frac{3}{4} \left( 1 + \frac{3}{2} \frac{a_e}{A_e} \right) Li_{5/2} \left( - e^{-a_e} \right) + a_e \left( 1 + \frac{3}{8} \frac{a_e}{A_e} \right) \nonumber \\
  && \left. \times Li_{3/2}  \left( - e^{-a_e} \right) + a_e^2  Li_{1/2}  \left( - e^{-a_e} \right)\right\} ,
  \label{deltaQJ_shot}
\end{eqnarray}
whereas the correlation between heat and current fluctuations is
\begin{eqnarray}
  && \langle  \delta \dot I_{J {\rm shot}} (\omega) \delta \dot Q^*_{J {\rm shot}} (\omega) \rangle
  \approx \frac{2}{e R_T} \int_\Delta^\infty \frac{\epsilon (\epsilon - eV)}{\sqrt{\epsilon^2 - \Delta^2}} \frac{d\epsilon}{e^{\beta_e (\epsilon - eV)} + 1}
  \nonumber \\
  %
  %
  &&\approx - \frac{2 (k_B T)^{3/2}}{e R_T} \sqrt{\frac{\pi \Delta}{2}} \left\{ \frac{9}{16 A_e} Li_{5/2} \left( - e^{-a_e} \right) \right. \nonumber \\
  && \left. + \frac{1}{2} \left( 1 + \frac{3}{4} \frac{a_e}{A_e} \right) Li_{3/2} \left( - e^{-a_e} \right) + a_e Li_{1/2}  \left( - e^{-a_e} \right) \right\} .
  \label{deltaIJdeltaQJ_shot}
\end{eqnarray}
Now we calculate the partial derivatives:
\begin{subequations} \label{derivs_IQ}
\begin{eqnarray}
  \frac{\partial I_J}{\partial (k_BT_e)} &\approx& \frac{\beta_e^2}{e R_T} \int_\Delta^\infty \frac{\epsilon (\epsilon - eV)}{\sqrt{\epsilon^2-\Delta^2}} \frac{e^{\beta_e(\epsilon - eV)}}{[e^{\beta_e(\epsilon - eV)} + 1]^2} d \epsilon \nonumber \\
  &\approx& - \frac{1}{e R_T} \sqrt{\frac{\pi \Delta}{2 k_BT}} \left\{ \frac{9}{16 A_e} Li_{3/2}\left( -e^{-a_e} \right) \right. \nonumber \\
  && + \frac{1}{2} \left( 1 + \frac{3}{4} \frac{a_e}{A_e} \right) Li_{1/2} \left( -e^{-a_e} \right) \nonumber \\
  && \left. + a_e Li_{- 1/2} \left( -e^{-a_e} \right) \right\} ,
  \label{dIJ_dTe}
\end{eqnarray}
\begin{eqnarray}
  \frac{\partial I_J}{\partial (eV)} &\approx& \frac{\beta_e}{e R_T} \int_\Delta^\infty \frac{\epsilon}{\sqrt{\epsilon^2-\Delta^2}} \frac{e^{\beta_e(\epsilon - eV)}}{[e^{\beta_e(\epsilon - eV)} + 1]^2} d \epsilon \nonumber \\
  &\approx& - \frac{1}{e R_T} \sqrt{\frac{\pi \Delta}{2 k_BT}} \left\{ \frac{3}{8 A_e} Li_{1/2}\left( -e^{-a_e} \right) \right. \nonumber \\
  && \left. + Li_{- 1/2} \left( -e^{-a_e} \right) \right\} ,
  \label{dIJ_deV}
\end{eqnarray}
\begin{eqnarray}
  \frac{\partial \dot Q_J}{\partial (k_BT_e)} &\approx& \frac{\beta_e^2}{e^2 R_T} \int_\Delta^\infty \frac{\epsilon (\epsilon-eV)^2}{\sqrt{\epsilon^2-\Delta^2}} \frac{e^{\beta_e(\epsilon - eV)} \, d \epsilon}{[e^{\beta_e(\epsilon - eV)} + 1]^2} ,
  \nonumber \\
  && \approx - \frac{k_BT_e}{e^2 R_T} \sqrt{\frac{\pi A_e}{2}} \left[ \frac{3}{4} \left( 1 + \frac{3}{2} \frac{a_e}{A_e} \right) Li_{3/2} \left( -e^{-a_e} \right) \right. \nonumber \\
  && + \left( a_e + \frac{3}{8} \frac{a_e^2}{A_e} \right) Li_{1/2} \left( -e^{-a_e} \right)
  \label{dQJ_dTe} \\
  && \left. + a_e^2 Li_{-1/2} \left( -e^{-a_e} \right) + \frac{45}{32 A_e} Li_{5/2} \left( -e^{-a_e} \right) \right] , \nonumber
\end{eqnarray}
and
\begin{eqnarray}
  \frac{\partial \dot Q_J}{\partial (eV)} &\approx& \frac{\beta_e}{e^2 R_T} \int_\Delta^\infty \frac{\epsilon (\epsilon-eV)}{\sqrt{\epsilon^2-\Delta^2}} \frac{e^{\beta_e(\epsilon - eV)} \, d \epsilon}{[e^{\beta_e(\epsilon - eV)} + 1]^2} \nonumber \\
  &=& k_BT_e \frac{\partial I_J}{\partial (k_BT_e)} . \label{dQJ_deV}
\end{eqnarray}
\end{subequations}
In Eqs.~(\ref{derivs_IQ}) we used the general derivation rules for polylogarithms to define
\begin{equation}
  Li_{- 1/2} \left( -e^{-a_e} \right) \equiv - \frac{d Li_{1/2} \left( -e^{-a_e} \right)}{d a_e} . \label{deriv_polylog}
\end{equation}

\begin{figure}[t]
  \centering
  \includegraphics[width=9cm]{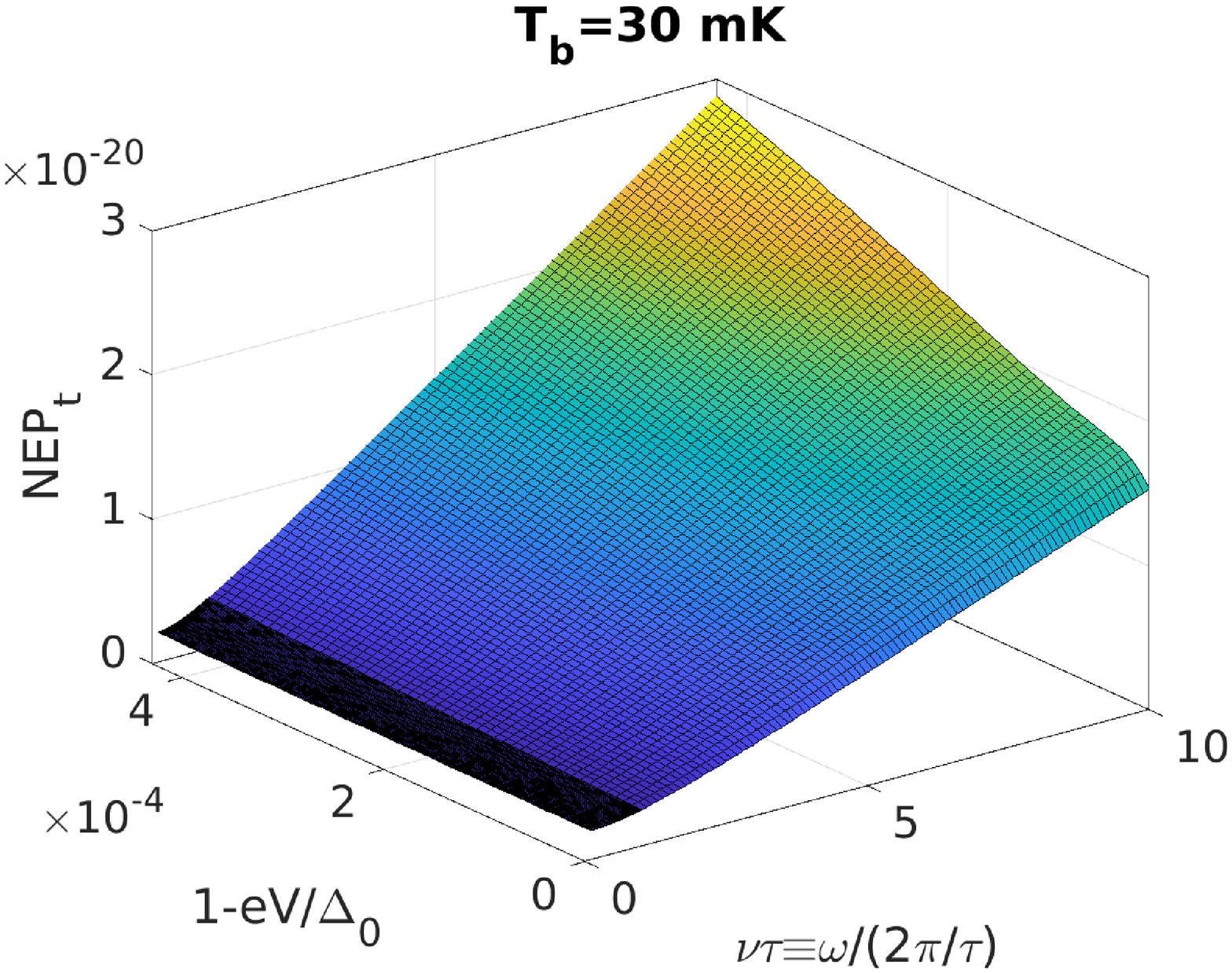}
  \includegraphics[width=9cm]{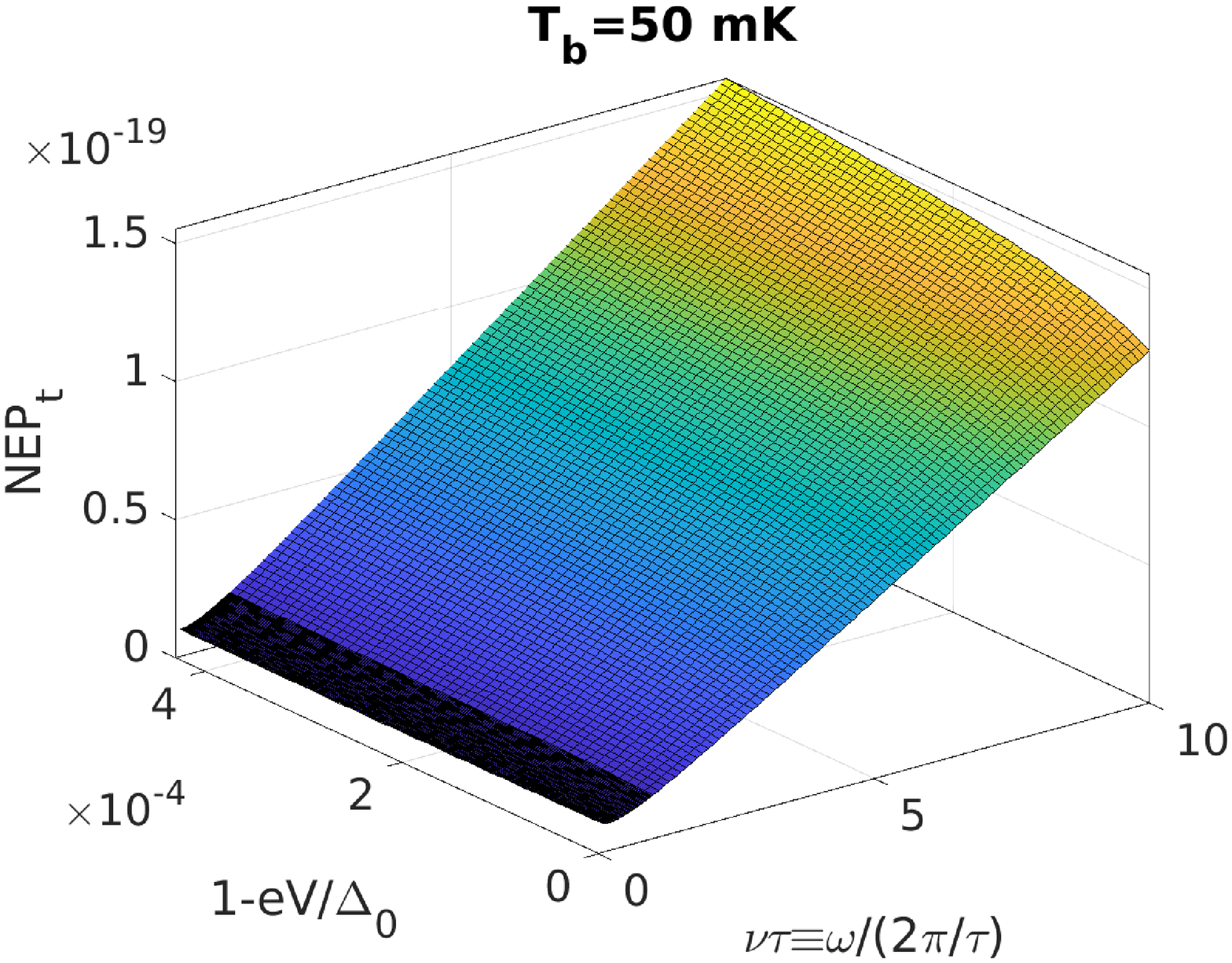}
  \caption{The ${\rm NEP}_t$ (without taking into consideration $\langle |\delta \dot{Q}_{oe\,{\rm shot}}(\omega)|^2 \rangle$) vs $\omega/(2\pi/\tau) \equiv \nu \tau$ and $1-eV/\Delta_0$ for two values of the bath temperature: $T_b = 30$~mK and $T_b = 50$~mK, as indicated.}
  \label{NEP_total}
\end{figure}

Using Eqs.~(\ref{NEP_tot})-(\ref{derivs_IQ}) we can calculate ${\rm NEP}_t$, which is plotted in Fig.~\ref{NEP_total}, for $\omega$ taking values in a relevant interval, $\omega \in [2\pi/(10 \tau), 2\pi/(\tau/10)]$, and for $T_b = 30$~mK and 50~mK.

\begin{figure}[t]
  \centering
  \includegraphics[width=9cm,bb=0 0 699 572]{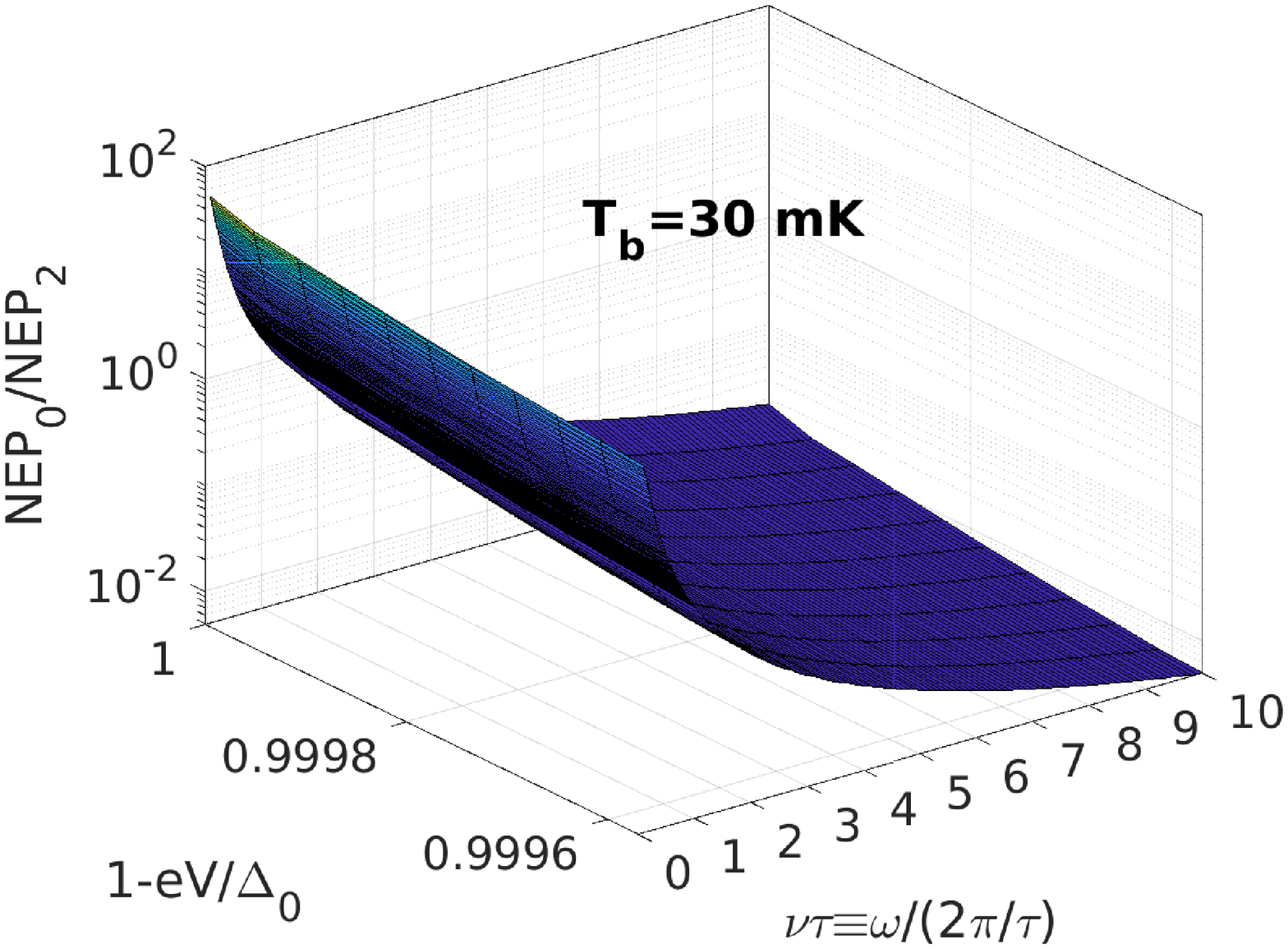}
  \includegraphics[width=9cm,bb=0 0 699 572]{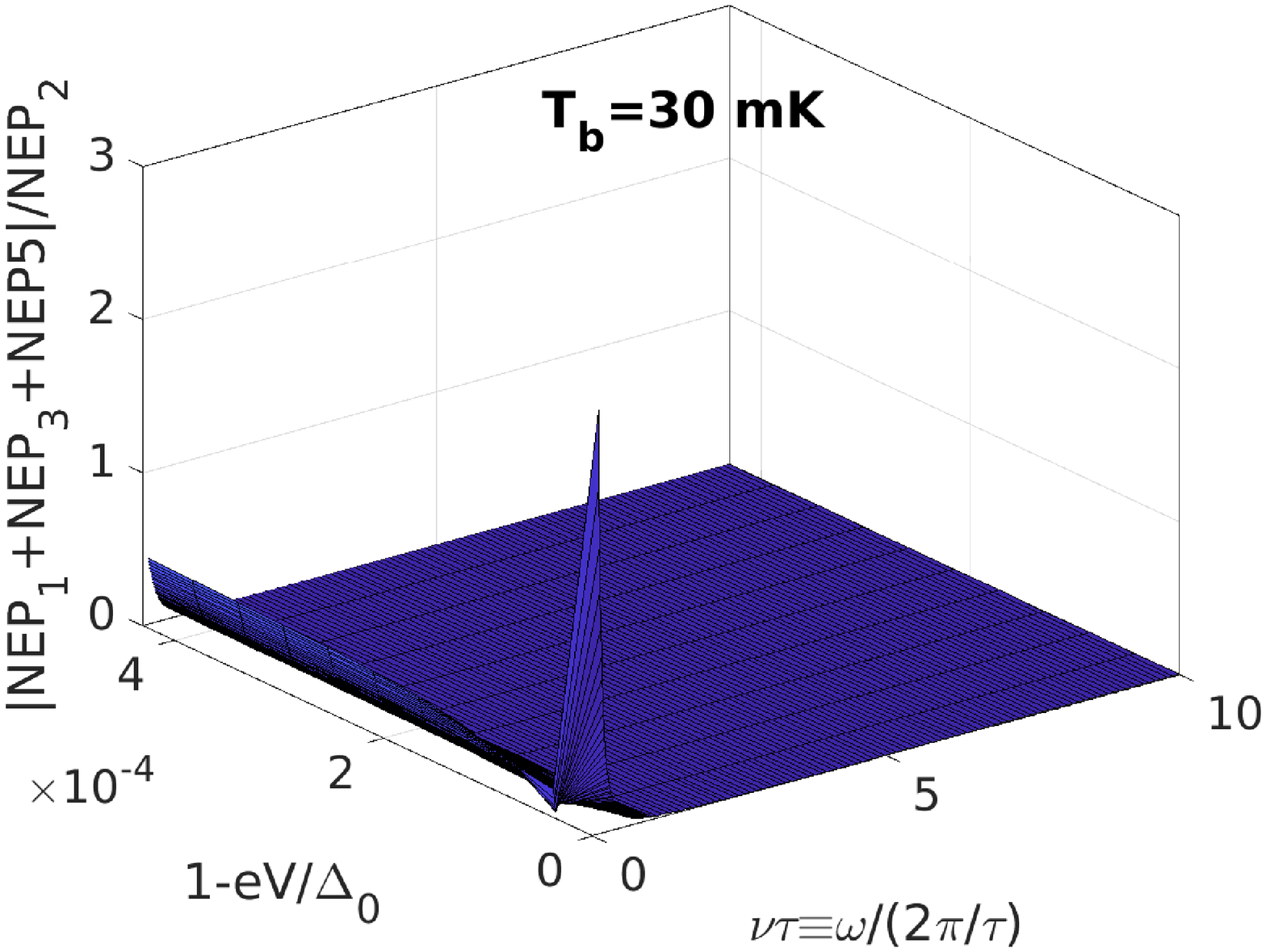}
  \caption{The ratios NEP$_0$/NEP$_2$ and $|{\rm NEP}_1 + {\rm NEP}_3 + {\rm NEP}_5|/{\rm NEP}_2$, for two values of $T_b$ -- NEP$_0$, NEP$_1$, NEP$_2$, NEP$_3$ and NEP$_5$ are the only relevant contributions to NEP$_t$. We observe that in the most part of the parameters range NEP$_2$ is the most dominant term.}
  \label{NEPterms_T2_T0}
\end{figure}

Among the terms NEP$_i$ ($i=0,\ldots,8$), the dominant one in most of the frequency range of interest is NEP$_2$.
In Fig.~\ref{NEPterms_T2_T0} we compare NEP$_2$ with NEP$_0$ and with the absolute value $|{\rm NEP}_1 + {\rm NEP}_3 + {\rm NEP}_5|$, for $T_b = 30$~mK -- for $T_b$ between 30 and 50~mK, the plots are rather similar.
All the other terms, namely NEP$_4$, NEP$_6$, NEP$_7$, and NEP$_8$, are over twenty orders of magnitude smaller than ${\rm NEP}_0 + {\rm NEP}_1 + {\rm NEP}_2 + {\rm NEP}_3 + {\rm NEP}_5$, so we shall neglect them in all the calculations.
The term $\langle |\delta \dot{Q}_{oe\,{\rm shot}}(\omega)|^2 \rangle$ is relevant only for integrating detectors, so we shall disregard it also in our analysis of SPCs.
Therefore, the relevant terms may be grouped in three categories, according to their $\omega$ dependence:
\begin{subequations} \label{NEP_cat}
\begin{equation}
  {\rm NEP}_t^2 \equiv A +  \frac{B}{|S_I(\omega,V)|^2} + \frac{C}{\omega^2 |S_I(\omega,V)|^2}
  \label{cat_tot}
\end{equation}
where
\begin{eqnarray}
  A &\equiv& \langle |\delta \dot{Q}_{ep\,{\rm shot}}(\omega)|^2 \rangle
  + 2 \langle |\delta\dot{Q}_{J {\rm shot}}(\omega)|^2\rangle \nonumber \\
  && - \frac{4 \langle \delta I_{J {\rm shot}}(\omega) \delta \dot{Q}^*_{J {\rm shot}}(\omega) \rangle }{ S_I(0,V) } , \label{cat_A} \\
  B &\equiv& 2 \langle |\delta I_{J {\rm shot}}(\omega)|^2\rangle , \label{cat_B}
\end{eqnarray}
and
\begin{eqnarray}
  C &\equiv& \frac{2}{e^2} \left(\frac{\partial\epsilon_F}{\partial N} \frac{\partial I_J}{\partial (eV)}\right)^2 \langle |\delta I_{J {\rm shot}}(\omega)|^2 \rangle
  \label{cat_C}
\end{eqnarray}
\end{subequations}

Vice-versa, from Eqs.~(\ref{NEP_cat}) and the definition of NEP, we get the expression for the current fluctuation,
\begin{equation}
  \langle |\delta I_J(\omega)|^2 \rangle = |S_I(\omega,V)|^2 {\rm NEP}_t^2
  = |S_I(\omega,V)|^2 A + B + \frac{C}{\omega^2} .
  \label{fluct_I_fin}
\end{equation}

\subsection{Single-photon counters} \label{subsec_SPC}

The calculation of the NEP is just a prerequisite and cannot be directly compared to the signal produced by the photon absorption in a SPC~\cite{ApplPhysLett.82.293.2003.Anghel, NatPhotonics.3.696.2009.Hadfield, ApplPhysLett.96.083505.2010.Santavicca, PhysRevApplied.3.014007.2015.Gasparinetti, PhysRevB.98.205414.2018.Brange}.
For example, Hadfield proposes two different definitions for the NEP~\cite{NatPhotonics.3.696.2009.Hadfield},
\begin{equation}
  {\rm NEP}_{H1} \equiv (h \nu / \eta) \sqrt{2D}
  \ {\rm and} \
  {\rm NEP}_{H2} \equiv \eta / (D \Delta t) , \label{Hadfield_NEP}
\end{equation}
to qualitatively adapt it to the SPC.
In Eqs.~(\ref{Hadfield_NEP}), $h\nu$ is the energy of the photon, $\eta$ is the detection efficiency (number of detected photons divided by the number of incident photons), $D$ is the dark count rate, and $\Delta t$ is the timing jitter of the detector (the variation of the time interval between the absorption of a photon and the generation of the output electrical pulse) -- notice that while ${\rm NEP}_{H1}$ has units of WHz$^{-1/2}$, ${\rm NEP}_{H2}$ is dimensionless.
While both definitions~(\ref{Hadfield_NEP}) represent qualitative scales for the performance of the detectors, they are of no practical use for our SPC, since they cannot discern between the cases when the photon can or cannot be observed.

Another estimation of the energy resolution of the SPC was used in~\cite{PhysRevApplied.3.014007.2015.Gasparinetti}, as $\delta E_G = {\rm NEP}_G \tau^{-1/2}$, where ${\rm NEP}_G \equiv {\rm NET}_G C_V$, is the noise equivalent power, ${\rm NET}_G$ is the noise equivalent temperature, and $\tau$ is the re-equilibration time of the detector.
This formula for the energy resolution represents the contribution of the NEP in a frequency window equal to $\tau^{-1}$.
But, in Section~\ref{sec_noise} we showed that in order to detect the photon, the current signal produced by its absorption should be bigger than the fluctuation of the current in the frequency window of the filter.
Using a too low frequency cutoff (like $1/\tau$, in this case), would drastically reduce not only the noise, but also (and especially) the pulse height -- as seen, for example, in Eqs.~(\ref{def_tau_c}) and (\ref{delta_IJ_V_m}) -- and therefore would decrease the detection capabilities of the SPC (in Fig.~\ref{time_evol_IJ_eV} we used a cutoff time $\tau_c = 1$~ns, for a time constant $\tau$ more than an order of magnitude bigger).
To put it simply, using ${\rm NEP}_G$ as an estimation of the noise in SPC gives a unrealistic (too optimistic) estimation of the detector's capabilities.

The importance of recognizing the signal produced by the photon (in the measured quantity) from the detector noise was shown theoretically in~\cite{ApplPhysLett.82.293.2003.Anghel} and was concretely emphasized by Santavicca et al. in Ref.~\cite{ApplPhysLett.96.083505.2010.Santavicca}.
In accordance with this, Brange et al.~\cite{PhysRevB.98.205414.2018.Brange} theoretically compare the temperature spike, produced by the photon absorption, with the Monte Carlo simulation of the noise in the detector.
Nevertheless, in both Refs.~\cite{ApplPhysLett.96.083505.2010.Santavicca} and \cite{PhysRevB.98.205414.2018.Brange} there are not provided concrete criteria for the possibility of detecting the photon -- for example, in Ref.~\cite{PhysRevB.98.205414.2018.Brange}, only the \textit{temperature variation} is \textit{visually} compared with the \textit{temperature noise}.
Therefore, in addition to this, here and in Ref.~\cite{ApplPhysLett.82.293.2003.Anghel}, we propose a concrete calculation method to compare the measured signal with the average fluctuation of the measured quantity in the relevant time scale of the experiment.


\end{document}